\documentclass[twocolumn,prb,aps,showpacs,floatfix,superscriptaddress,preprintnumbers,amsmath,amssymb]{revtex4}

\usepackage{graphicx}

\begin{document}

\newcommand{\mfs}{Mn$_{1-x}$Fe$_{x}$Si}
\newcommand{\mcs}{Mn$_{1-x}$Co$_{x}$Si}
\newcommand{\fcs}{Fe$_{1-x}$Co$_{x}$Si}

\newcommand{\mb}{$\mu_0\,M/B$}
\newcommand{\dmdb}{$\mu_0\,\mathrm{d}M/\mathrm{d}B$}
\newcommand{\ddmddb}{$\mathrm{\mu_0\Delta}M/\mathrm{\Delta}B$}
\newcommand{\cm}{$\chi_{\rm M}$}
\newcommand{\cac}{$\chi_{\rm ac}$}
\newcommand{\rechi}{${\rm Re}\,\chi_{\rm ac}$}
\newcommand{\imchi}{${\rm Im}\,\chi_{\rm ac}$}

\newcommand{\ozz}{$\langle100\rangle$}
\newcommand{\ooz}{$\langle110\rangle$}
\newcommand{\ooo}{$\langle111\rangle$}
\newcommand{\too}{$\langle211\rangle$}

\newcommand{\ea}{\textit{et al.}}

\title{Magnetic phase diagram of MnSi inferred from magnetization and ac susceptibility}

\author{A. Bauer}
\affiliation{Technische Universit\"at M\"unchen, Physik-Department E21, D-85748 Garching, Germany}

\author{C. Pfleiderer}
\affiliation{Technische Universit\"at M\"unchen, Physik-Department E21, D-85748 Garching, Germany}

\date{\today}

\begin{abstract}
We report simultaneous measurements of the magnetization and the ac susceptibility across the magnetic phase diagram of single-crystal MnSi. In our study we explore the importance of the excitation frequency, excitation amplitude, sample shape, and crystallographic orientation. The susceptibility, {\dmdb}, calculated from the magnetization, is dominated by pronounced maxima at the transition from the helical to the conical and the conical to the skyrmion lattice phase. The maxima in {\dmdb} are not tracked by the ac susceptibility, which in addition varies sensitively with the excitation amplitude and frequency at the transition from the conical to the skyrmion lattice phase. The same differences between {\dmdb} and the ac susceptibility exist for {\mfs} ($x=0.04$) and {\fcs} ($x=0.20$). Taken together our study establishes consistently for all major crystallographic directions the existence of a single pocket of the skyrmion lattice phase in MnSi, suggestive of a universal characteristic of all B20 transition metal compounds with helimagnetic order.
\end{abstract}


\vskip2pc

\maketitle

\section{Introduction}
B20 transition metal compounds such as MnSi, {\fcs}, or FeGe exhibit long-wavelength helimagnetic order at low temperatures. As emphasized already in Landau-Lifshitz, Vol.\,VIII, Sec. 52, \cite{Landau} the helimagnetic order in B20 compounds reflects a well defined set of hierarchical energy scales. On the strongest scale ferromagnetic exchange interactions favor parallel spin-alignment. Due to the lack of inversion symmetry of the B20 crystal structure this is followed on intermediate scales by Dzyaloshinsky-Moriya (DM) spin-orbit interactions, favoring perpendicular spin alignment. The combination of ferromagnetic exchange and DM interactions generates a long wavelength helical modulation. The chirality of the DM interaction and thus of the helical modulation is thereby determined by the enantiomorph of the B20 structure. Finally, on the weakest energy scale higher order spin-orbit coupling, also referred to as crystal electric field effects, aligns the propagation direction of the helical modulations along either {\ooo} or {\ozz} depending on the system. 

Recently, interest in the spin order in B20 transition metal compounds has revived, when a small phase pocket just below the helimagnetic transition temperature $T_c$, long known as the A phase, was identified as the first example of a skyrmion lattice in a magnetic material. \cite{Muehlbauer:Science2009,Muenzer:PRB2010,Pfleiderer:JPCM2010} The skyrmion lattice represents a novel form of magnetic order composed of spin vortices with a non-zero topological winding number. The skyrmion lattice is thereby always perpendicular to the applied field; i.e., the pinning with respect to the crystal lattice is extremely weak. \cite{Muehlbauer:Science2009,Adams:preprint2012} Perhaps most remarkable, the coupling of electric currents to the skyrmion lattice is extremely efficient, leading to emergent electrodynamics and sizable spin torque effects at ultra-low current densities. \cite{Jonietz:Science2010,Schulz:NaturePhysics2012}

The microscopic identification of the skyrmion lattice was initially based on small-angle neutron scattering (SANS) in MnSi, {\mfs}, {\mcs}, and {\fcs}. \cite{Muehlbauer:Science2009,Muenzer:PRB2010,Pfleiderer:JPCM2010} The non-zero topological winding number, as the defining aspect of the skyrmion lattice, was first inferred from topological contributions to the Hall effect \cite{Neubauer:PRL2009} and recently confirmed in a rather sophisticated small-angle neutron scattering study. \cite{Adams:PRL2011} Real-space imaging by means of Lorentz force microscopy confirmed the formation of the skyrmion lattice in thin samples of selected B20 transition metal compounds, notably {\fcs}, FeGe, and MnSi. \cite{Yu:Nature2010,Yu:NatureMaterials2011, Tonumura:NanoLetters2012} As the most recent development a skyrmion lattice has been observed in the multiferroic compound Cu$_2$OSeO$_3$, \cite{Seki:Science2012,Adams:PRL2012} which crystallizes in the same $P2_{1}3$ space group as the B20 compounds. 

The large body of macroscopic and microscopic information on the magnetic phase diagram and skyrmion lattice phase in B20 transition metal bulk compounds available so far has been explained in terms of Ginzburg-Landau theory taking into account thermal Gaussian fluctuations. \cite{Muehlbauer:Science2009} For the thin samples studied by Lorentz force microscopy the effects of reduced spatial extent were considered, as they destabilize the conical phase, thereby supporting other routes to skyrmion lattices. \cite{Han:PRB2010} Neither the data of microscopic properties nor these theoretical models suggest the existence of a more complex phase diagram with multiple phases.

In contrast, thirty years ago Kadowaki {\ea} reported that the A phase in MnSi consists of two phase pockets, \cite{Kadowaki:JPSJ82} putatively suggesting several phases with different spin structures of nontrivial topology. The evidence for the two phase pockets was thereby purely based on the presence of two kinks in the temperature dependence of the magnetization when traversing the A phase near its upper and lower boundary. In contrast, no kinks were observed in the magnetization for a field in the center of the A phase, suggesting the absence of an A phase. However, Kadowaki {\ea} recorded the magnetization with a Faraday balance which requires a field gradient. In turn, the magnetization data shown by Kadowaki {\ea} may have been subject to parasitic magnetic torques and did not represent the longitudinal magnetization. Moreover, as pointed out by Kadowaki {\ea} the temperature control of their setup was rather unsatisfactory. Data were therefore recorded by means of temperature sweeps. This may have caused small systematic temperature differences between different temperature sweeps. Finally, magnetoresistance data as a function of magnetic field, also reported by Kadowaki {\ea}, suggested the presence of only a single pocket of the A phase. 

The question of several phase pockets was recently followed up in a study of FeGe, where an even more complex magnetic phase diagram including claims of mesophases with liquid-crystal-like appearance were proposed. \cite{Wilhelm:PRL2011} Yet, the phase diagram in FeGe was purely inferred from the ac susceptibility as measured at a unique excitation frequency of 1\,kHz in a Quantum Design PPMS. Moreover, the FeGe sample studied, being vapor transport grown, had an irregular shape causing ill-defined demagnetizing fields and a poor residual resistivity ratio characteristic for defects and disorder (see also the note added at the end of this paper concerning Cu$_{2}$OSeO$_{3}$). 

The possible existence of multiple pockets in the magnetic phase diagram of MnSi and FeGe has been explained in terms of a Landau theory, where the ground state is stabilized by a combination of magnetic anisotropies and a controversial gradient term favoring soft amplitude fluctuations. \cite{Wilhelm:PRL2011,Leonov:preprint2010} The possible existence of a more complex magnetic phase diagram is therefore believed to provide putative evidence for a different microscopic mechanism at the heart of the skyrmion lattice than that considered so far.

To clarify the magnetic phase diagram in MnSi and other B20 transition metal compounds we report in the following simultaneous measurements of the magnetization and ac susceptibility in high-quality single-crystal MnSi. In our study we carefully explored the importance of the excitation frequency and amplitude for the ac susceptibility, as well as the sample shape and crystallographic orientation. As the skyrmion lattice phase is embedded in a spin-flop phase, also known as conical phase, we covered the entire magnetic phase diagram. We finally also performed exploratory measurements in {\mfs} and {\fcs} to demonstrate the generic nature of our observations for other B20 transition metal compounds and in the presence of disorder and defects.

In our study of MnSi we establish consistently for all major crystallographic directions the existence of a single pocket of the A phase. In contrast, there is no evidence whatsoever pointing at any additional subphases. The most prominent aspect of our results concerns the observation of maxima in the susceptibility {\dmdb}, calculated from the magnetization, at the border between the helical, the conical and the skyrmion lattice phase. General thermodynamic considerations and the imaginary part of the ac susceptibility establish, that the maxima in {\dmdb} are the characteristic of processes on extremely slow time-scales involving very large magnetic objects, consistent with a crossover or slightly broadened first-order transitions at the helical to conical transition or the boundary of the A phase, respectively. Surprisingly, the maxima in {\dmdb} are not tracked by the real part of the ac susceptibility, {\rechi}, which even varies sensitively with the excitation amplitude and frequency at the transition to the skyrmion lattice. Reproducing the temperature dependence of the magnetization reported by Kadowaki {\ea} (Ref.\,\cite{Kadowaki:JPSJ82}), we are finally able to show that the magnetic phase diagram reported in Ref.\,\cite{Kadowaki:JPSJ82} is not correct.  

Moreover, our results question the evidence for the rather complex magnetic phase diagram of FeGe reported by Wilhelm {\ea}, \cite{Wilhelm:PRL2011} which was purely based on the ac susceptibility recorded at a single frequency of 1\,kHz. Taken together with data we recorded for {\mfs} and {\fcs}, our study suggests that the magnetic phase diagram we observe in MnSi represents a universal characteristic of all B20 transition metal compounds with helimagnetic order.


\section{Experimental Methods}
\label{methods}

For our studies of the magnetic phase diagram of MnSi a large single crystal was grown in a bespoke ultrahigh vacuum compatible image furnace. \cite{Neubauer:RSI2011} The feed rods and seed rods used for the float zoning were prepared from high-purity elements and cast in a UHV-compatible radio-frequency induction furnace equipped with a bespoke all-metal-sealed Huykin crucible and casting mold. \cite{Neubauer:preprint2012} Samples were oriented by means of Laue x-ray diffraction and cut with a wire saw. The sample surfaces were gently polished. The residual resistivity ratio (RRR) of samples prepared from the same ingot was very high exceeding 300, characteristic of an excellent sample quality. For the preliminary studies in {\mfs} and {\fcs} reported at the end of this paper single-crystal samples were grown following the same procedure as for MnSi. Previous studies on the same {\mfs} and {\fcs} single crystals were reported in Refs.\,\cite{Bauer:PRB2010,Muenzer:PRB2010}.

\begin{table*}
\centering
\caption{
Samples investigated in this study. Magnetic fields were applied along the direction labeled (a).
\label{samples}
}
\begin{tabular}{llllllllll}
\hline\noalign{\smallskip}
\hline\noalign{\smallskip}
system & sample & shape & (a) size, orientation & (b) size, orientation & (c) size, orientation & $N$\\
\hline\noalign{\smallskip}
MnSi & 1 & bar & 6\,mm, {\ozz} &  1\,mm, {\ooz} &  1\,mm, {\ooz} & 0.074\\
MnSi & 2 & cube & 1\,mm, {\ozz} &  1\,mm, {\ooz} &  1\,mm, {\ooz} & 0.333\\
MnSi & 3 & cube & 1\,mm, {\ooz} &  1\,mm, {\ooo} &  1\,mm, {\too} & 0.333\\
MnSi & 4 & platelet & 0.115\,mm, {\ozz} &  3\,mm, {\ooz} &  1\,mm, {\ooz} & 0.837\\
{\mfs}, $x=0.04$ & 5 & bar & 6\,mm, {\ozz} &  1\,mm, {\ooz} &  1\,mm, {\ooz} & 0.074\\
{\fcs}, $x=0.20$ & 6 & bar & 6\,mm, {\ozz} &  1\,mm, {\ooz} &  1\,mm, {\ooz} & 0.074\\
\noalign{\smallskip}\hline
\noalign{\smallskip}\hline
\end{tabular}\\
\end{table*}

To track the importance of demagnetizing fields and crystallographic orientation on the magnetic phase diagram of MnSi four samples with three different sample geometries were studied as summarized in Table\,\ref{samples}. The sample shapes investigated consisted of a bar, two cubes, and a thin platelet. Most of our measurements were carried out on the bar-shaped single-crystal denoted as sample 1. The dimensions of sample 1 were $6\times1\times1\,{\rm mm^3}$ with the long direction parallel to a {\ozz} axis. The small faces of sample 1 were perpendicular to {\ooz} axes. The magnetic field was applied along the long direction of sample 1 to minimize the effects of demagnetizing fields. The samples of {\mfs} and {\fcs}, denoted as samples 5 and 6 in Table\,\ref{samples}, had the same shape and orientation as sample 1.

Samples 2 and 3 were cubic and served to study the role of the crystallographic orientation under unchanged sample geometry. Sample 2 was oriented such that two opposing faces were oriented perpendicular to a {\ozz} axis. The other four faces were oriented perpendicular to a {\ooz} axes. This contrasts sample 3, which was oriented such that the faces were perpendicular to either {\ooo}, {\ooz}, or {\too}, respectively. Samples 2 and 3 allowed us to reproduce data for $\langle110\rangle$ and by rotating the cube through 90$^{\circ}$ to record data for field parallel to a $\langle100\rangle$ or $\langle111\rangle$ under unchanged sample geometry. Sample 4 was a thin platelet of $3\times1\times0.115\,{\rm mm^3}$ (length$\times$width$\times$thickness); i.e., the dimensions were characteristic  of samples used in magnetotransport measurements. The smallest direction of the platelet was thereby parallel to a $\langle100\rangle$ axis, while the long direction, equivalent to the current direction in transport measurements, was parallel to a $\langle110\rangle$ axis. 

The demagnetizing factors of the samples were determined by approximating the sample shape as a rectangular prism. For the field directions given above they are $N_1=0.074$, $N_{2}=N_{3}=0.333$, and $N_4=0.837$ for samples 1, 2, 3, and 4, respectively. As summarized below all temperature and field dependencies were found to be qualitatively independent of the sample geometry. Demagnetizing fields only affected the absolute values of the transition fields in excellent agreement with the calculated demagnetization factors given above. In addition, the quantitative size of the anomalies at the phase boundaries were found to decrease with increasing demagnetizing factor while the range of the inconsistencies increased. For lack of space, all experimental data are shown as a function of applied field without correction for demagnetizing fields, while the phase diagrams inferred from the data are shown as a function of both applied and internal field, respectively.

The magnetization and the ac susceptibility were measured in a Quantum Design physical properties measurement system (PPMS). Data points were recorded at fixed temperature and magnetic field values. Unlike common practice in many studies no data were recorded while the field or the temperature were swept continuously. At each given temperature and field value the magnetization was determined by means of a standard extraction method. This was followed by the measurement of the ac susceptibility. The ac field was only turned on while recording the susceptibility data. We note that all data reported here were recorded using the same sample thermometer. However, in comparison to other studies reported in the literature there may be small systematic differences of the absolute temperature values on the order of a few \% (see also the clarification of Ref.\,\cite{Muehlbauer:Science2009}). The ac susceptibility was recorded in the frequency range from 10\,Hz to 10\,kHz. For typical values of the resistivity and the susceptibility in the regime of the A phase of MnSi, $\rho_{\textrm{MnSi}}(28\,\textrm{K}) \approx 35\,\mu\Omega\textrm{cm}$, $\chi_{\rm MnSi}\approx 0.3$, and the skin depth $\delta_{\textrm{MnSi}}(28\,\textrm{K})=  (2\rho/(\omega \mu_{0}(\chi + 1)))^{1/2} \approx 0.26 \cdot f^{-1/2} \textrm{ms}^{-1/2}$. In the range of excitation frequencies from 10\,Hz to 10\,kHz applied in our study the skin depth was between 82\,mm and 2.6\,mm, respectively, and thus considerably larger than the relevant sample dimensions. As the resistivity of {\mfs} and {\fcs} is larger than that of MnSi the same conclusion is true for these compounds.

In the course of our studies we became aware of two minor technical limitations of the PPMS  as concerns simultaneous measurements of the ac susceptibility and magnetization. First, when repeating measurements of the ac susceptibility on the same sample and sample support (remounting the sample from scratch between measurements), the signal may include a small temperature-independent signal contribution that varies slightly between different measurements. Second, the  ac susceptibility measured for parameters where highly reversible behavior is expected, i.e., at high fields and/or high temperatures,  differs by several \% from the susceptibility, {\dmdb}, calculated from the magnetic field dependence of the magnetization. 

We have therefore recalibrated the measured ac susceptibility of all samples and sample orientations with respect to the susceptibility {\dmdb} calculated from the magnetization, by the same procedure as follows. We subtracted \textit{the same constant value} of all ac susceptibility data, $\Delta\chi=0.025$, at all temperatures and fields. We then multiplied \textit{all} of our ac susceptibility data by a factor of 1.1. This way the ac susceptibility and values of {\dmdb} agreed for field and temperature values well away from the magnetically ordered state, notably at high magnetic fields and/or high temperatures. We believe that the small deviation visible in Fig.\,\ref{FSLayoutSusVgl2}\,(a) reflects this mismatch. Taken together, the conclusions of the work reported here are independent of this recalibration of the ac susceptibility data. 

\section{Experimental results}

The presentation of our experimental results is organized as follows. We begin with the magnetic field dependence of sample 1, where the ac susceptibility data were recorded at a low excitation amplitude of $0.1\,{\rm mT}$ and an excitation frequency of 10\,Hz unless explicitly stated differently. This is followed by a comparison of typical changes of the ac susceptibility for larger excitation amplitudes, notably $0.3\,{\rm mT}$ and  $1\,{\rm mT}$ at 10\,Hz, and higher excitation frequencies of 91\,Hz and 911\,Hz at 0.1\,mT. 

In the second part of this section we present the temperature dependence of the magnetization and ac susceptibility of sample 1 for $0.1\,{\rm mT}$ and 10\,Hz, followed by data illustrating the role of changes of excitation amplitudes and frequencies. This way we establish the consistency of the rather unusual temperature dependence with the magnetic field dependence. In particular, we are able to show the consistency of the evidence for a single A phase seen in the magnetic field dependence with the data reported by Kadowaki {\ea}. \cite{Kadowaki:JPSJ82}

The third part of the experimental results section concerns variations of the ac susceptibility under changes of sample shape and crystallographic orientation. On the one hand this allows us to point out potential pitfalls when inferring the magnetic phase diagram of B20 compounds from ac susceptibility data alone. On the other hand our studies reveal a regime of inconsistency between magnetization and ac susceptibility at the boundary of the A phase. The results section concludes with a presentation of preliminary results on {\mfs} and {\fcs}.

\begin{figure}
\includegraphics[width=0.35\textwidth]{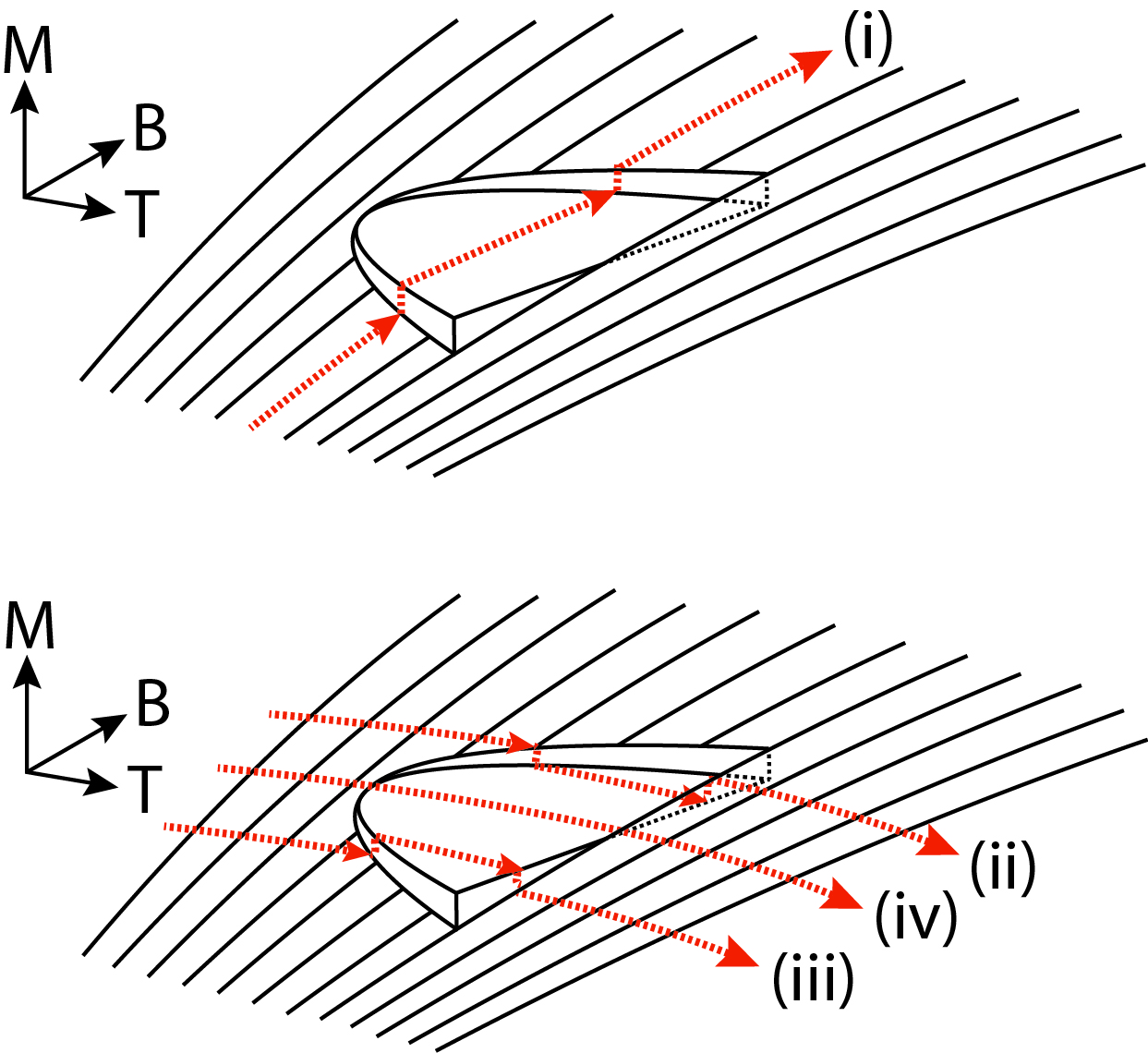}
\caption{(Color online) Schematic depiction of the magnetization as a function of temperature and magnetic field. In the regime of the skyrmion lattice phase the magnetization is characterized by a tilted plateaux. The boundary of this plateaux is a discontinuous (first order) step. The traces illustrate typical scans as a function of magnetic field or temperature. In the experimental data the discontinuous steps are smeared out.}
\label{APhaseschematic}
\end{figure}

Given the amount and complexity of the data presented in the following it is helpful to present at first in Fig.\,\ref{APhaseschematic} a qualitative summary as a guide of the essential features of the magnetization in the regime of the skyrmion lattice phase. The key characteristic of the magnetization in the skyrmion lattice phase is a plateau, that is slightly tilted with respect to the surrounding phases. At the boundary of this plateau the magnetization changes discontinuously (first order). This discontinuous step is smeared out in the experimental data. As a function of increasing magnetic field, depicted by trace (i), the magnetization shows always two first-order transitions at which the magnetization increases. However, as a function of increasing temperature more complex behavior is observed. For a temperature scan in the vicinity of the upper boundary of the skyrmion lattice phase, depicted by trace (ii), the magnetization displays a decreasing step followed by an increasing step. In contrast, in a temperature scan in the vicinity of the lower boundary, depicted by trace (iii), the magnetization shows an increasing step followed by a decreasing step. Finally, for an intermediate field it is, in principle, possible to obtain temperature scans that show zero or one transitions, cf. trace (iv). This is a purely accidental effect, which accounts for our data as well as that reported by Kadowaki {\ea}. \cite{Kadowaki:JPSJ82} Yet, behavior such as that depicted by trace (iv) is clearly no evidence of the existence of several phase pockets.

As part of the presentation of our experimental results we define various characteristic transition and crossover fields and temperatures. All definitions are thereby based on {\dmdb} and {\ddmddb} with corresponding features in {\rechi} and {\imchi}. Characteristic field values were determined under increasing magnetic fields. Unfortunately even for a minimalist description it is necessary to account for four transitions as a function of field (helical $\to$ conical $\to$ skyrmion lattice $\to$ conical $\to$ ferromagnetic) and three transitions as a function of temperature (helical $\to$ paramagnetic; conical $\to$ skyrmion lattice $\to$ paramagnetic). When trying to track the width of the transitions this requires a large number of characteristic fields and temperatures. Even though it might appears tedious, we feel that this is necessary to demonstrate the full consistency between field and temperature dependent data. We thereby note, that some of the definitions given are rather empirical and cannot be followed in the data very accurately. For clarity we use a short-hand notation to mark all characteristic field and temperature values in the figures. A summary of how the field and temperature values are defined is shown in Tables \ref{definitions-1} and \ref{definitions-2}. More detailed definitions are given in the text, when the experimental evidence for the various transitions is presented first. As mentioned above all data are presented as measured experimentally, i.e., as a function of applied magnetic field. The magnetic phase diagrams are shown as a function of applied magnetic field as well as a function of internal magnetic field, where the effects of demagnetizing fields have been corrected. 

%

\begin{table*}
\centering
\caption{
Summary of characteristic transition and cross-over fields and associated labels used in the figures shown in this paper. The signatures in {\dmdb} were used to define the actual field values; these correspond within the accuracy of our data to the signatures in {\rechi} and {\imchi} listed in this table. The definition of the signatures refers to data recorded for increasing magnetic field. Note that our definition of $B_{\rm A1}^{-}$ and $B_{\rm A2}^{+}$ corresponds to values referred to as $B_{\rm A1}$ and $B_{\rm A2}$ in the literature.
\label{definitions-1}
}
\begin{tabular}{llllllllll}
\hline\noalign{\smallskip}
\hline\noalign{\smallskip}
in text & in figures & signature in {\dmdb} & signature in {\rechi} & signature in {\imchi}\\
\hline\noalign{\smallskip}
$B_{\rm c1}^{-}$ & (1) & begin {\dmdb}$\neq${\rechi} & begin {\dmdb}$\neq${\rechi} & no signature\\
$B_{\rm c1}$ & (2) & peak & point of inflection & peak for large amplitude\\
$B_{\rm c1}^{+}$ & (3) & end {\dmdb}$\neq${\rechi} & end {\dmdb}$\neq${\rechi} & no signature \\ 
$B_{\rm A1}^{-}$ & (4) & begin of peak & begin of deviation from & begin peak\\
&& & {\rechi} in conical phase &\\
$ B_{\rm A1}^{+}$ & (5) & begin of plateau of reduced & begin of plateau of reduced & end peak\\
& & {\dmdb} in A phase & {\rechi} in A phase &\\
$B_{\rm A2}^{-}$ & (6) & end of plateau of reduced & end of plateau of reduced & begin peak\\
& & {\dmdb} in A phase & {\rechi} in A phase &\\
$B_{\rm A2}^{+}$ & (7) & end of peak & end of deviation from & end peak \\
&& & {\rechi} in conical phase &\\
$B_{\rm c2}$ & (8) & point of inflection & point of inflection & no signature\\
\noalign{\smallskip}\hline
\noalign{\smallskip}\hline
\end{tabular}\\
\vspace{8mm}
\centering
\caption{
Summary of characteristic transition and cross-over temperatures and associated labels used in the figures shown in this paper. The signatures in {\ddmddb} were used to define the actual temperature values; these correspond within the accuracy of our data to the signatures in {\rechi} and {\imchi} listed in this table. The definition of the signatures refers to data recorded for increasing temperature. Note that our definition of $T_{\rm A1}^{-}$ and $T_{\rm A2}^{+}$ corresponds to values referred to as $T_{\rm A1}$ and $T_{\rm A2}$ in the literature.
\label{definitions-2}
}
\begin{tabular}{llllllllll}
\hline\noalign{\smallskip}
in text & in figures & signature in {\ddmddb} & signature in {\rechi} & signature in {\imchi}\\
\hline\noalign{\smallskip}
$T_{\rm A1}^{-}$ & (i) &  begin of maximum &  begin of deviation from & begin peak\\
&&&{\rechi} in conical phase& \\
$T_{\rm A1}^{+}$ & (ii) &  end of maximum & end {\ddmddb}$\neq${\rechi} & end peak\\
$T_{\rm A2}^{+}$ &  (iii) & end of deviation from &  end of deviation from & no signature\\
&&{\ddmddb} in conical phase&{\rechi} in conical phase& \\
$T_{\rm c2}$ & (iv) & point of inflection & point of inflection & no signature\\
\noalign{\smallskip}\hline
\noalign{\smallskip}\hline
\end{tabular}\\
\end{table*}

\subsection{Magnetic field dependence}
\label{field}

\begin{figure}
\includegraphics[width=0.45\textwidth]{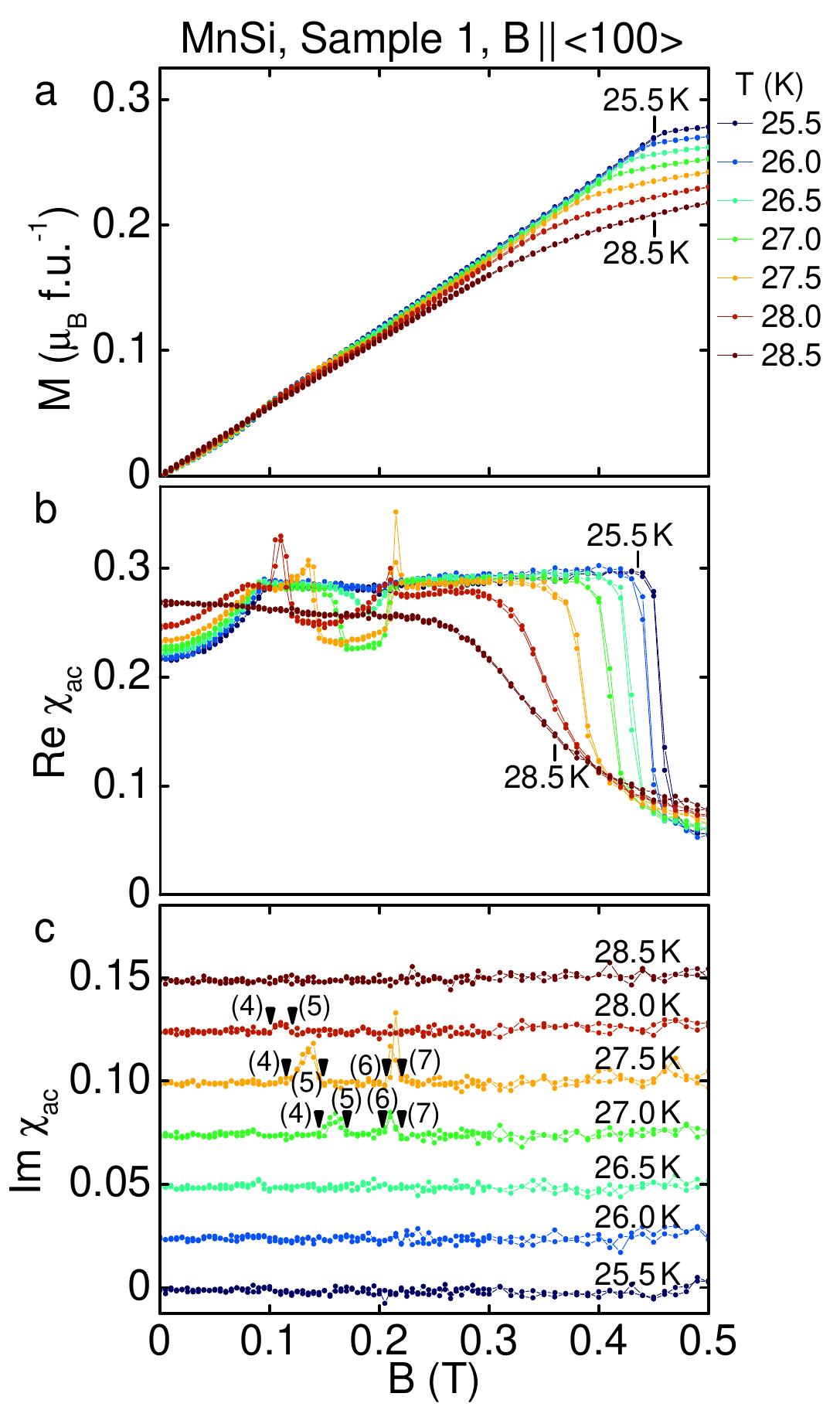}
\caption{(Color online) Typical magnetization and ac susceptibility data of sample 1 as a function of applied magnetic field. Data were recorded for various temperatures in the regime of the A phase. The ac susceptibility data were recorded for an excitation amplitude of 0.1\,mT parallel to $\langle100\rangle$ at 10\,Hz. See also Fig.\,\ref{FSLayoutSusVgl2} for definitions of the transition fields. (a) Magnetization as a function of applied magnetic field. (b) Real part of the ac susceptibility as a function of applied magnetic field. (c) Imaginary part of the ac susceptibility as a function of applied magnetic field. Data shown in panel (c) have been shifted by constant value of 0.025 for clarity.}
\label{FSLayout}
\end{figure}

\begin{figure*}
\includegraphics[width=1.0\textwidth]{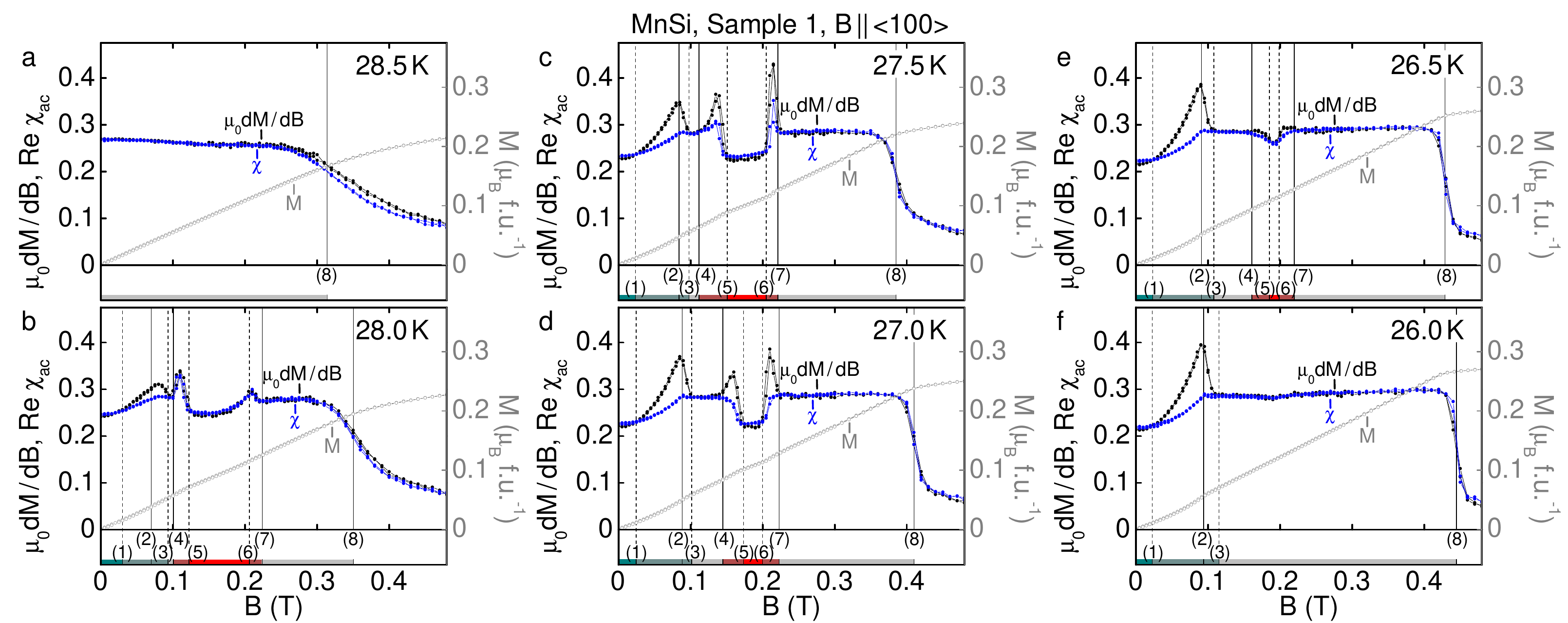}
\caption{(Color online) Comparison of the magnetic field dependence of the magnetization, $M$, the susceptibility calculated from the magnetization, {\dmdb}, and the real part of the ac susceptibility, {\rechi}. Data shown here were recorded in sample 1 for various temperatures in the regime of the A phase. The ac susceptibility data were recorded for an excitation amplitude of 0.1\,mT along {\ozz} at 10\,Hz.}
\label{FSLayoutSusVgl2}
\end{figure*}

Shown in Fig.\,\ref{FSLayout}\,(a) is the magnetization $M$ of sample 1 as a function of magnetic field along {\ozz} for temperatures in the vicinity of the helimagnetic transition temperature $T_c$. We begin with an overview and an account of the global features and return to a detailed comparison including the definition of the transition fields in Fig.\,\ref{FSLayoutSusVgl2}. Data for both increasing and decreasing fields are shown, where the field sweeps were started at $-1\,{\rm T}$ and $+1\,{\rm T}$ for increasing and decreasing fields, respectively. For this field history data correspond to the field-cooled behavior in temperature scans presented below. None of the field sweeps displayed hysteresis between increasing and decreasing field at the resolution of the methods used. All our data are perfectly consistent with previous studies. 

Starting from $B=0$ the magnetization increases almost linearly. Well below $T_c$ the initial quasi-linear increase for $B<B_{c1}$ arises from an increasing anharmonicity of the helical modulation, a slight reorientation of the helical modulation, and changes of domain population as seen in neutron scattering studies; see, e.g., Refs.\,\cite{Lebech:1993,Grigoriev:PRB2006a,Grigoriev:PRB2006b}. Depending on the field direction the reorientation at $B_{c1}$ represents a cross-over or a symmetry breaking second order phase transition. In contrast, for $B>B_{c1}$ the linear increase is a characteristic of a closing-in of the spin-flop phase (the conical phase) up to $B_{c2}$, the transition to the field-polarized ferromagnetic state. On the field scale shown in Fig.\,\ref{FSLayout} the magnetization appears almost saturated above $B_{c2}$. However, as reported in the literature up to $\sim$33\,T, \cite{Sakakibara:JPSJ1982} the highest fields studied, the magnetization is unsaturated and nonlinear characteristic of itinerant magnetism. On the field scale shown here, the rather small changes of the magnetization in the A phase are not obvious.

Deviations from the quasilinear field dependence of the magnetization are best seen in the susceptibility. The real part of the ac susceptibility, {\rechi}, is shown in Fig.\,\ref{FSLayout}\,(b) for the same field and temperature values as the magnetization shown in Fig.\,\ref{FSLayout}\,(a). At low temperatures and low fields the ac susceptibility increases gently before reaching a well-defined plateau above $B_{c1}$. Again data for increasing and decreasing fields are shown, where we find no hysteresis between increasing and decreasing fields.

For temperatures just below $T_c$ a reduced value of the susceptibility is observed for magnetic fields between $\sim100\,{\rm mT}$ and $\sim200\,{\rm mT}$. Previous studies have established that this corresponds to the A phase. \cite{Gregory:JMMM1992,Thessieu:JPCM1997} In other words the magnetization in the A phase is characterized by a plateau of reduced slope as a function of increasing field, which implies that there must be a regime of increased slope connecting the conical phase with the A phase at its boundaries. Indeed, as shown in Fig.\,\ref{FSLayout}\,(b) for an excitation amplitude of 0.1\,mT and excitation frequency of 10\,Hz chosen here, {\rechi} displays narrow maxima at the transition between the conical phase and the A phase.

The imaginary part of the ac susceptibility, {\imchi}, shown in Fig.\,\ref{FSLayout}\,(c), shows clear maxima at the transition fields between the conical phase and the A phase [note that data shown in Fig.\,\ref{FSLayout}\,(c) have been shifted by a constant value of 0.025 for clarity]. This provides evidence that the narrow maxima in {\rechi} at the border of the A phase reflects the response of large magnetic objects on very slow time scales which originate in a slightly broadened first-order step in the magnetization, i.e., broadened $\delta$ anomalies in {\rechi}. The finite imaginary part of the ac susceptibility provides unambiguous evidence of the dissipation typically observed in regimes of very slow dissipative processes. This does not contrast the absence of hysteresis between increasing and decreasing fields within the resolution of the methods used, since the pinning may just be very weak.

Neither {\rechi} nor {\imchi} display any additional structure such as double peaks or additional shoulders at the phase boundary or deep inside the A phase. Thus there is no evidence that would hint at additional phase transitions. This is consistent with a wide range of microscopic studies we have performed (mostly neutron scattering), which do not provide any experimental evidence suggesting several pockets of the A phase either.

It is now instructive to compare in further detail the susceptibility calculated from the magnetization, {\dmdb}, shown in Fig.\,\ref{FSLayout}\,(a) with the real part of the ac susceptibility, {\rechi}, shown in Fig.\,\ref{FSLayout}\,(b). At $T=28.5\,{\rm K}$ shown in Fig.\,\ref{FSLayoutSusVgl2}\,(a), i.e., above the helimagnetic transition, {\dmdb} and {\rechi} agree essentially (we believe that the small difference of the data is due to a small systematic error as explained above). 

With decreasing temperature, shown in Figs.\,\ref{FSLayoutSusVgl2}\,(b) through \ref{FSLayoutSusVgl2}\,(f), we find in the regime of the helical to conical transition that {\dmdb} increases much faster than {\rechi} and displays a pronounced maximum and an abrupt drop. The absence of hysteresis between increasing and decreasing fields  is thereby characteristic of an extremely soft magnetic state in which the ac susceptibility, probing tiny hysteresis loops, does not fully track {\dmdb}. At the same time we find a tiny contribution to {\imchi} within our detection limit. This suggests that hysteretic losses are very small at the helimagnetic to conical transition and reveals a peculiar characteristic of the magnetic properties of MnSi, which, to the best of our knowledge, has not been noticed before. 

To be able to track the characteristics of the helical to conical transition we define three field values (see also Table\,\ref{definitions-1}). The lower field, $B_{\rm c1}^-$, denotes the point where {\dmdb} and {\rechi} begin to differ, the transition field $B_{\rm c1}$ is located at the maximum of {\dmdb}, and finally there is the upper field $B_{\rm c1}^+$, above which {\dmdb} and {\rechi} no longer differ. Values of $B_{\rm c1}$ agree thereby with the transition fields reported in the literature. The simplest explanation for the deviation between {\dmdb} and {\rechi} is the response of large magnetic objects at very slow time scales for instance due to a reorientation of the helical modulation.

We turn now to the A phase, where the same qualitative difference between {\dmdb} and {\rechi} emerges like that seen near $B_{\rm c1}$; see also Figs.\,\ref{FSLayoutSusVgl2}\,(b) through \ref{FSLayoutSusVgl2}\,(f). Here, clear maxima exist in {\dmdb}. Yet, these maxima are not tracked by {\rechi} which is always smaller than {\dmdb}. This similarity between the boundary of the A phase and the helical to conical transition, which arise microscopically on different ground, implies that the phenomenology in the magnetization and ac susceptibility by itself cannot be taken as evidence for the formation of mesophases of nontrivial spin topology.

Analogous to the helical to conical transition we define the following transition fields of the A phase. $B^-_{\rm A1}$ denotes the onset of the maximum in {\dmdb} at the transition to the A phase. For most of the parameter range studied this agrees with the field value where {\dmdb} and {\rechi} differ first. Likewise $B^+_{\rm A1}$ represents the upper field where the maximum in {\dmdb} is completed; in the ac susceptibility this corresponds to the point where {\dmdb} and {\rechi} merge and the plateau of the reduced ac susceptibility in the A phase begins. Analogous transition fields $B^-_{\rm A2}$ and $B^+_{\rm A2}$ are defined at the second maximum in {\dmdb} and the corresponding features in {\rechi}, respectively. By comparison with previous work $B^-_{\rm A1}$ corresponds to the low-field boundary reported in the literature, typically denoted as $B_{\rm A1}$, while $B^+_{\rm A2}$ corresponds to the high-field boundary of the A phase, typically denoted as $B_{\rm A2}$. 

\begin{figure}
\includegraphics[width=0.4\textwidth]{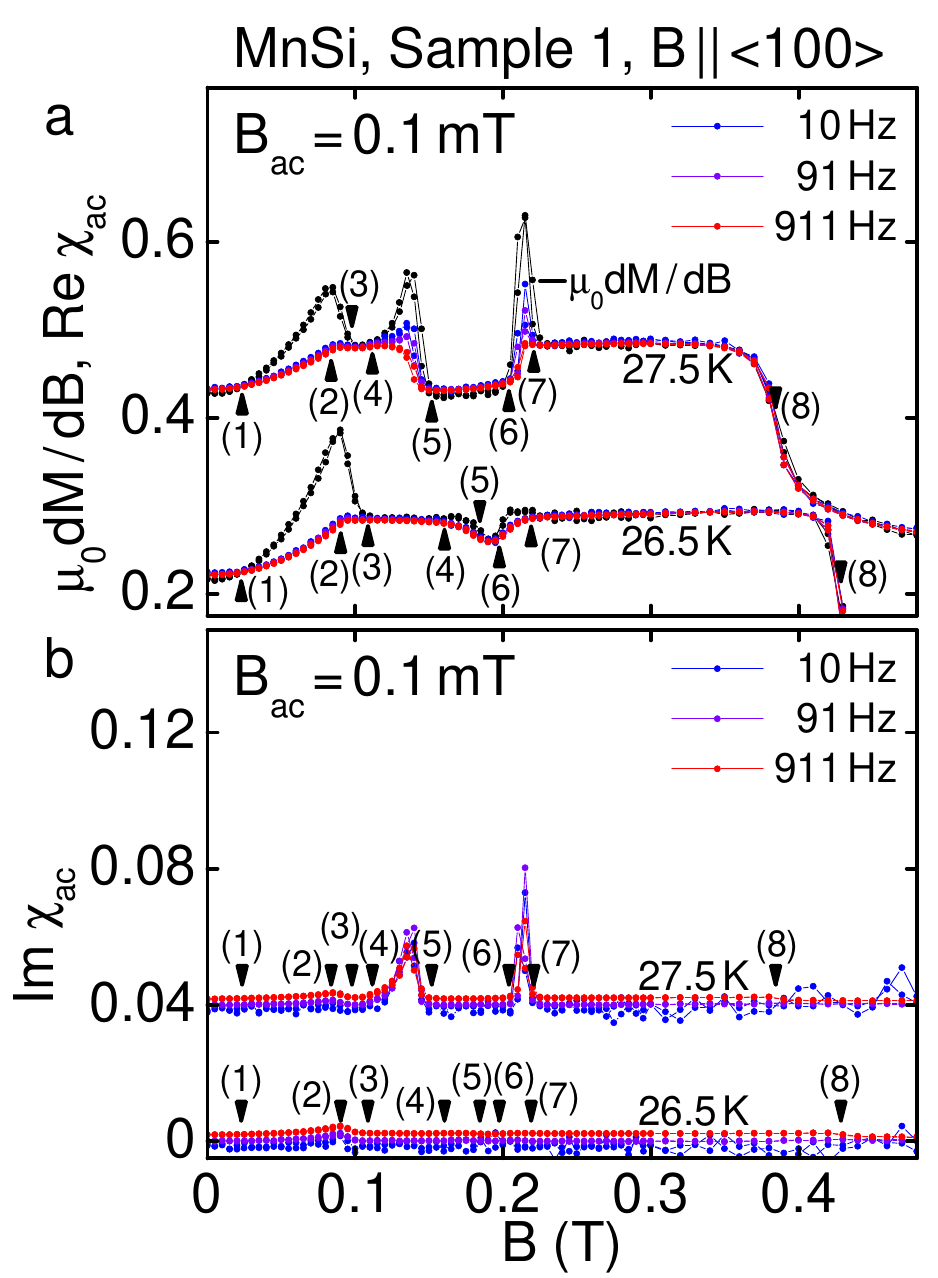}
\caption{(Color online) Typical magnetic field dependence of the susceptibility inferred from the magnetization, {\dmdb}, and the real and imaginary parts of the ac susceptibility, {\rechi} and {\imchi}, respectively,  at selected temperatures for various excitation frequencies. Data were recorded for an excitation amplitude of 0.1\,mT. Curves at different temperatures have been shifted by constant values for clarity. See Table \ref{definitions-1} for a summary of the definitions of characteristic field values.}
\label{FSLayoutFreq}
\end{figure}

\begin{figure}
\includegraphics[width=0.4\textwidth]{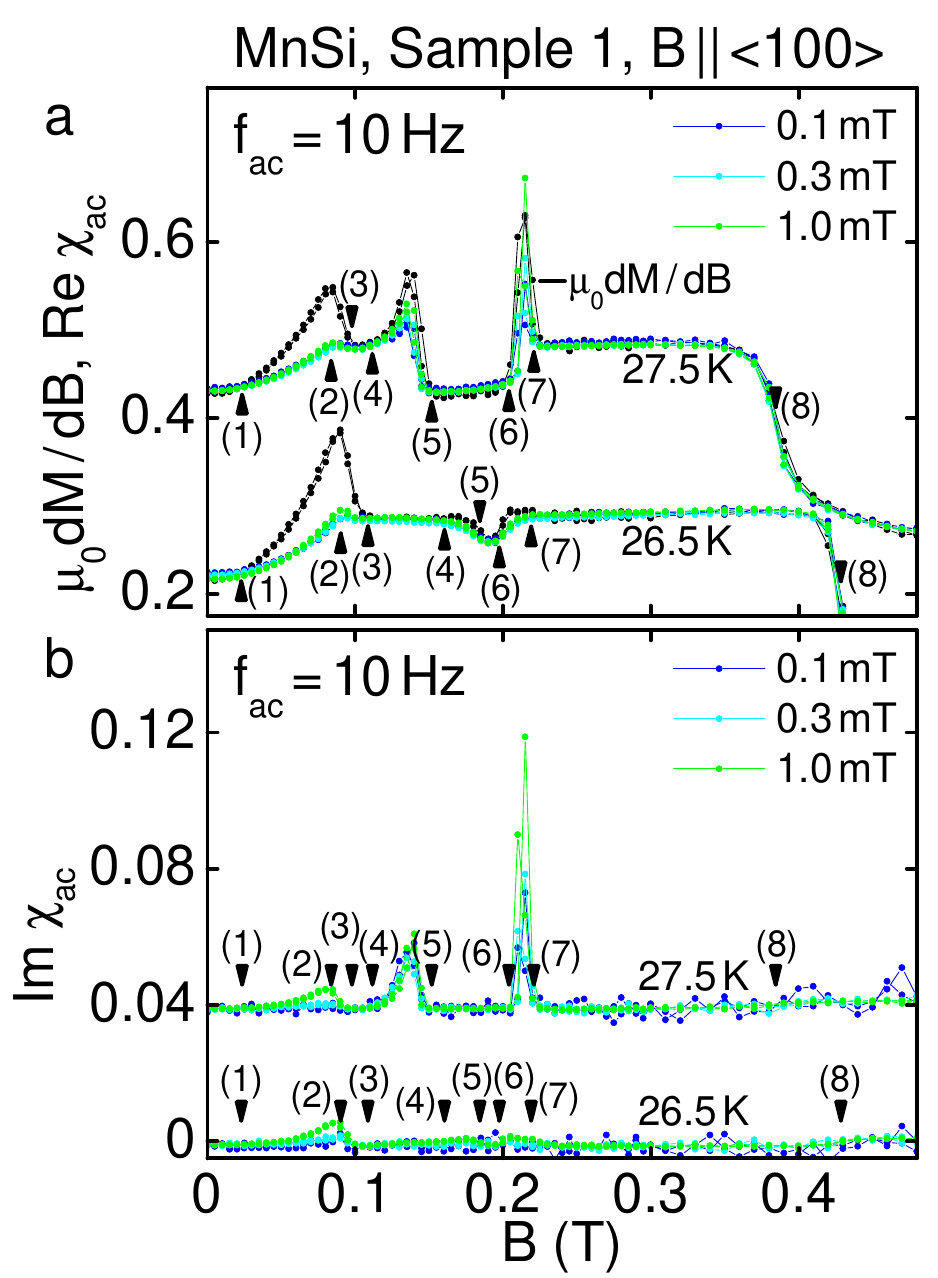}
\caption{(Color online) Typical magnetic field dependence of the susceptibility inferred from the magnetization, {\dmdb}, and the real and imaginary parts of the ac susceptibility, {\rechi} and {\imchi}, respectively,  at selected temperatures for various excitation amplitudes. Data were recorded for an excitation frequency of 10\,Hz. Curves at different temperatures have been shifted by constant values for clarity. See Table \ref{definitions-1} for a summary of the definitions of characteristic field values.}
\label{FSLayoutAmp}
\end{figure}

A key observation of our study concerns the absence of the narrow maxima in {\rechi} for sufficiently large excitation frequencies as compared with {\dmdb} for most of the temperatures studied. In fact, {\rechi} frequently displays a reduction, where the onset of the reduction agrees with $B^-_{\rm A1}$ in {\dmdb}. The differences between {\dmdb} and {\rechi} thereby depends on the precise choice of the excitation amplitude and frequency as illustrated in Figs.\,\ref{FSLayoutFreq} and \ref{FSLayoutAmp}. 

The importance of the excitation frequency is illustrated in Fig.\,\ref{FSLayoutFreq}\,(a), which displays data at two typical temperatures in the A phase for various frequencies and the same excitation amplitude of 0.1\,mT. Here the narrow maxima at the border of the A phase decrease with increasing excitation frequency. As emphasized above, this has nothing to do with the skin depth. For the standard excitation frequency of 911\,Hz, typically used by the PPMS, no maxima are observed. This compares with the amplitude dependence recorded for a fixed excitation frequency of 10\,Hz as shown in Figs.\,\ref{FSLayoutAmp}\,(a) and \ref{FSLayoutAmp}\,(b). Here the narrow maxima increase with increasing amplitude.

Even though the precise features in the ac susceptibility at the border of the A phase vary and depend on the measurement parameters, the values of  $B^-_{\rm A1}$ and $B^+_{\rm A2}$ are essentially unchanged; i.e., the outer boundary of the A phase in most studies reported in the literature were determined correctly. However, even though there is a pronounced frequency and amplitude dependence neither the real nor the imaginary part of the ac susceptibility displays additional structure such as a double peak or an additional shoulder at the boundary or deep inside the A phase that might hint at additional phase transitions. Instead they are characteristic of a small range with a response of large magnetic objects on slow time scales, consistentwith a slightly smeared first-order transition. Thus, the frequency and amplitude dependence underscore an inherent risk of inferring erroneous features of the phase diagram when the ac susceptibility is not measured carefully. 

\subsection{Temperature dependence}
\label{temperature}

We now turn to the temperature dependence of the magnetization and the ac susceptibility to establish the consistency with the field-dependent data reported above. This brings us also to the origin of the phase diagram reported by  Kadowaki {\ea}. \cite{Kadowaki:JPSJ82} It is thereby helpful to keep in mind the qualitative picture shown in Fig.\,\ref{APhaseschematic}, where the discontinuous steps in the the magnetization are smeared out in the experimental data. 

\begin{figure}
\includegraphics[width=0.45\textwidth]{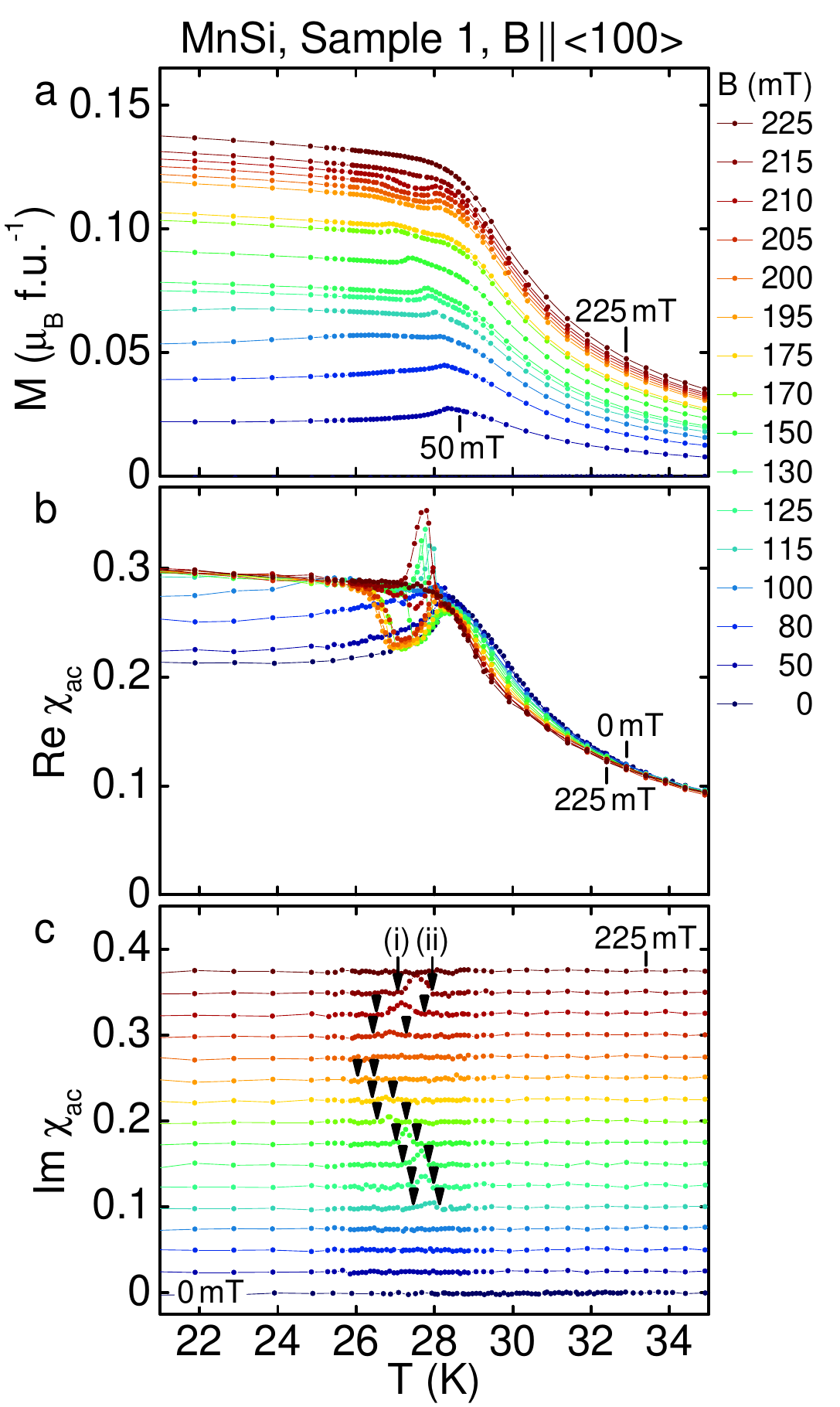}
\caption{(Color online) Typical temperature dependence of the magnetization and the ac susceptibility of sample 1 at various magnetic fields in the regime of the A phase. Data shown here were recorded in sample 1. The ac susceptibility data were recorded for an excitation amplitude of 0.1\,mT along {\ozz} at 10\,Hz. See Fig.\,\ref{TSLayoutSusVgl2} for definitions of the transition temperatures. (a) Magnetization as a function of temperature. (b) {\rechi} as a function of temperature. (c) {\imchi} as a function of temperature. Data in panel (c) have been shifted by constant values of 0.025 for clarity.}
\label{TSLayout}
\end{figure}

Shown in Fig.\,\ref{TSLayout}\,(a) is an overview of the temperature dependence of the magnetization of sample 1 in the vicinity of the helimagnetic to paramagnetic transition for various applied magnetic fields up to 225\,mT. The ac susceptibility data shown in Fig.\,\ref{TSLayout} were recorded for an excitation amplitude of 0.1\,mT at 10\,Hz along {\ozz}. In order to determine the temperature dependence of the susceptibility as inferred from the measured magnetization, we have measured $M(T)$ at field values that differed by 5\,mT and computed {\ddmddb} as a reasonable approximation of {\dmdb}.

We begin by describing the data qualitatively and return to detailed definitions of characteristic transition and cross-over temperatures further below. Again a shorthand notation is used for the characteristic temperatures, which is summarized in Table\,\ref{definitions-2}. 

With increasing temperature the magnetization initially varies only weakly. For temperatures $T>T_c$ a point of inflection marks the onset of a strong Curie Weiss dependence as a key characteristic of the paramagnetic state at high temperatures. Details of the transition region are best illustrated in the ac susceptibility as shown in Fig.\,\ref{TSLayout}\,(b). For instance, the point of inflection seen in the magnetization at a temperature slightly above the helimagnetic transition is also seen in {\rechi}. In finite magnetic fields this point of inflection remains unchanged for fields up to $\sim100\,{\rm mT}$ and is shifted to lower temperatures for slightly larger fields. As reported elsewhere, we have recently been able to show that the intermediate regime between the helimagnetic transition and the point of inflection (marked IM in Figs.\,\ref{PDLayoutBext} and \ref{PDLayoutBint}) exhibits all the characteristics of a Brazovskii transition \cite{Janoschek:2012} in contrast to earlier claims of a skyrmion liquid or Bak-Jensen transition. \cite{Roessler:Nature2006,Pappas:PRL2009,Grigoriev:PRB2010,Pappas:PRB2011}

Between $\sim115\,{\rm mT}$ and $\sim210\,{\rm mT}$, i.e., in the field range of the A phase, the most prominent feature of {\rechi} is a maximum along the upper and lower field boundary of the A phase that gives way to a minimum for intermediate fields. This is accompanied by a maximum in {\imchi} in a small temperature interval on the low-temperature side of the A phase as shown in Fig.\,\ref{TSLayout}\,(c) (for clarity curves have been shifted by a constant value of 0.025 in this panel). Marked by small arrows are in Fig.\,\ref{TSLayout}\,(c) the lower and upper boundary of the maximum in {\imchi}. As shown below this corresponds to the transition temperatures $T_{\rm A1}^-$ and $T_{\rm A1}^+$ in {\ddmddb}. 

\begin{figure*}
\includegraphics[width=1.0\textwidth]{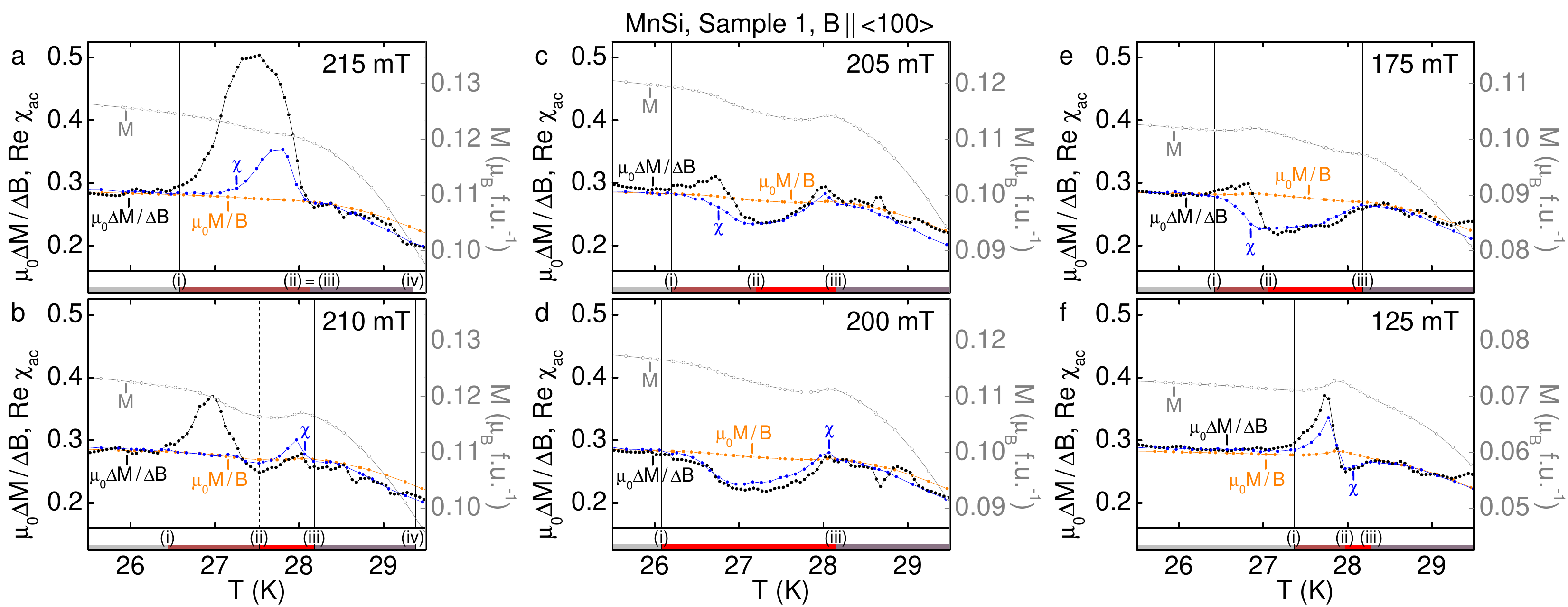}
\caption{(Color online) Comparison of the temperature dependence of the magnetization, $M$, the susceptibility inferred from the magnetization {\ddmddb}, and the real part of the ac susceptibility, {\rechi}. Also shown is the ratio {\mb}. Data shown here were recorded in sample 1 for various magnetic fields in the regime of the A phase. The ac susceptibility data were recorded for an excitation amplitude of 0.1\,mT along {\ozz} at 10\,Hz.}
\label{TSLayoutSusVgl2}
\end{figure*}

A detailed comparison of the temperature dependence of {\ddmddb} with {\rechi} at selected magnetic fields is shown in Fig.\,\ref{TSLayoutSusVgl2}. As in Fig.\,\ref{TSLayout} the ac susceptibility data were recorded simultaneously for an excitation frequency of 10\,Hz and an excitation amplitude of 0.1\,mT along {\ozz}. In the following we define various transition and cross-over temperatures. As for the magnetic field dependence a main observation in our study concerns the differences between {\rechi} and {\ddmddb}. 

Starting with the temperature dependence at 215\,mT, shown in Fig.\,\ref{TSLayoutSusVgl2}\,(a), the magnetization shows a shallow minimum that is accompanied by a pronounced maximum in {\ddmddb} in the same temperature range. This behavior corresponds to trace (ii) in Fig.\,\ref{APhaseschematic}, where the two transitions are smeared out and close together. The maximum in {\ddmddb} corresponds to the narrow maxima seen in the field sweeps at the upper boundary of the A phase. It extends over a wide temperature range, since the transition line of the A phase for this field value displays a very weak field dependence (cf. Fig.\,\ref{PDLayoutBext}). We define the transition temperatures $T_{\rm A1}^-$ and $T_{\rm A2}^+$ at the onset and the end of the maximum of {\ddmddb}, respectively, assuming that it consists essentially of two overlapping maxima. As summarized below these temperatures coincide perfectly with the phase boundary inferred from the field sweeps.

\begin{figure}
\includegraphics[width=0.4\textwidth]{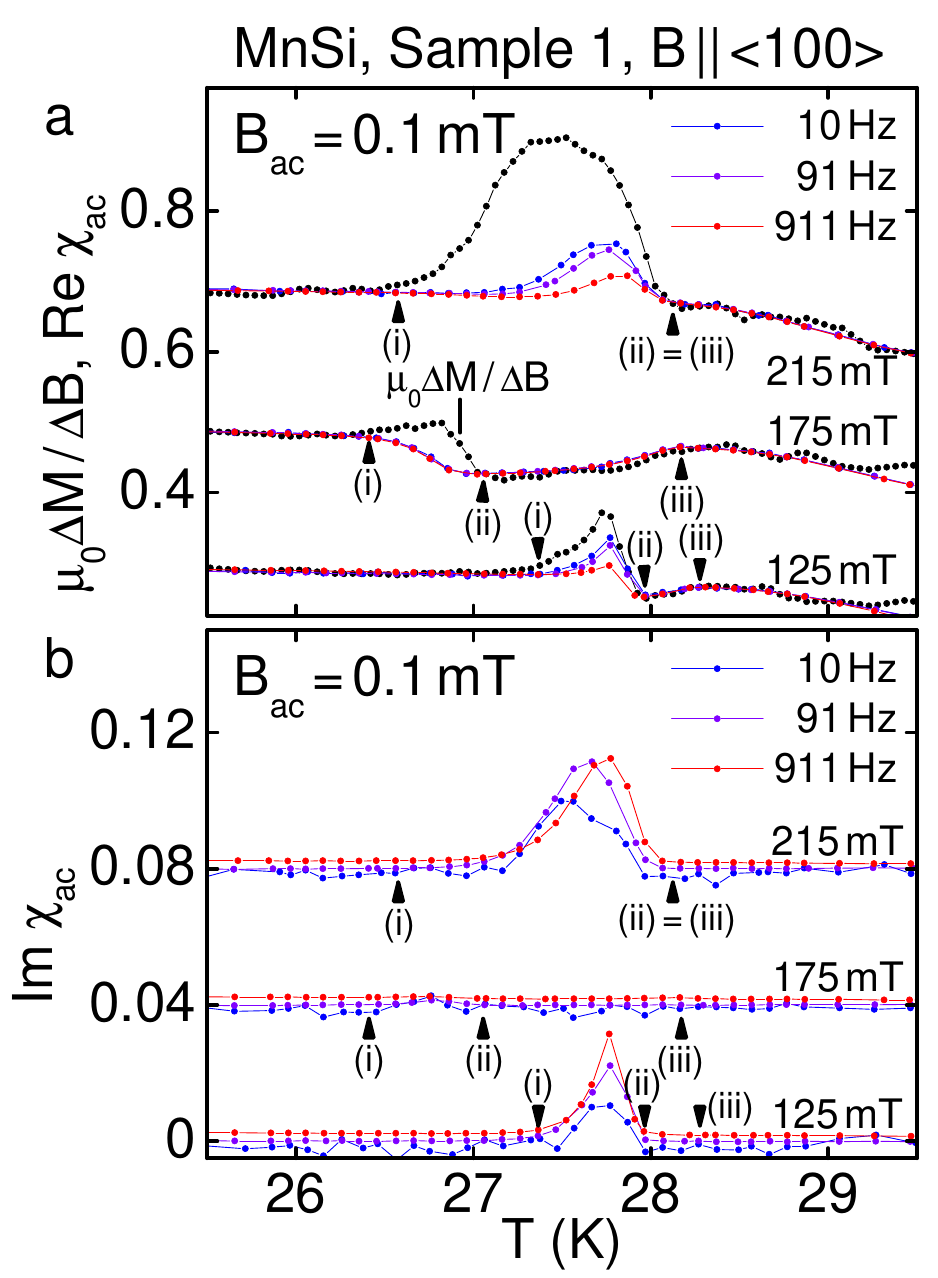}
\caption{(Color online) Typical temperature dependence of the susceptibility inferred from the magnetization, {\ddmddb}, and the real and imaginary parts of the ac susceptibility, {\rechi} and {\imchi}, respectively, at selected fields for various excitation frequencies. Data were recorded for an excitation amplitude of 0.1\,mT. {\ddmddb} was calculated from magnetization data recorded for a field difference of 5\,mT. Curves at different magnetic fields have been shifted by constant values for clarity. See Table \ref{definitions-2} for a summary of the definitions of temperature values.}
\label{TSLayoutFreq}
\end{figure}

\begin{figure}
\includegraphics[width=0.4\textwidth]{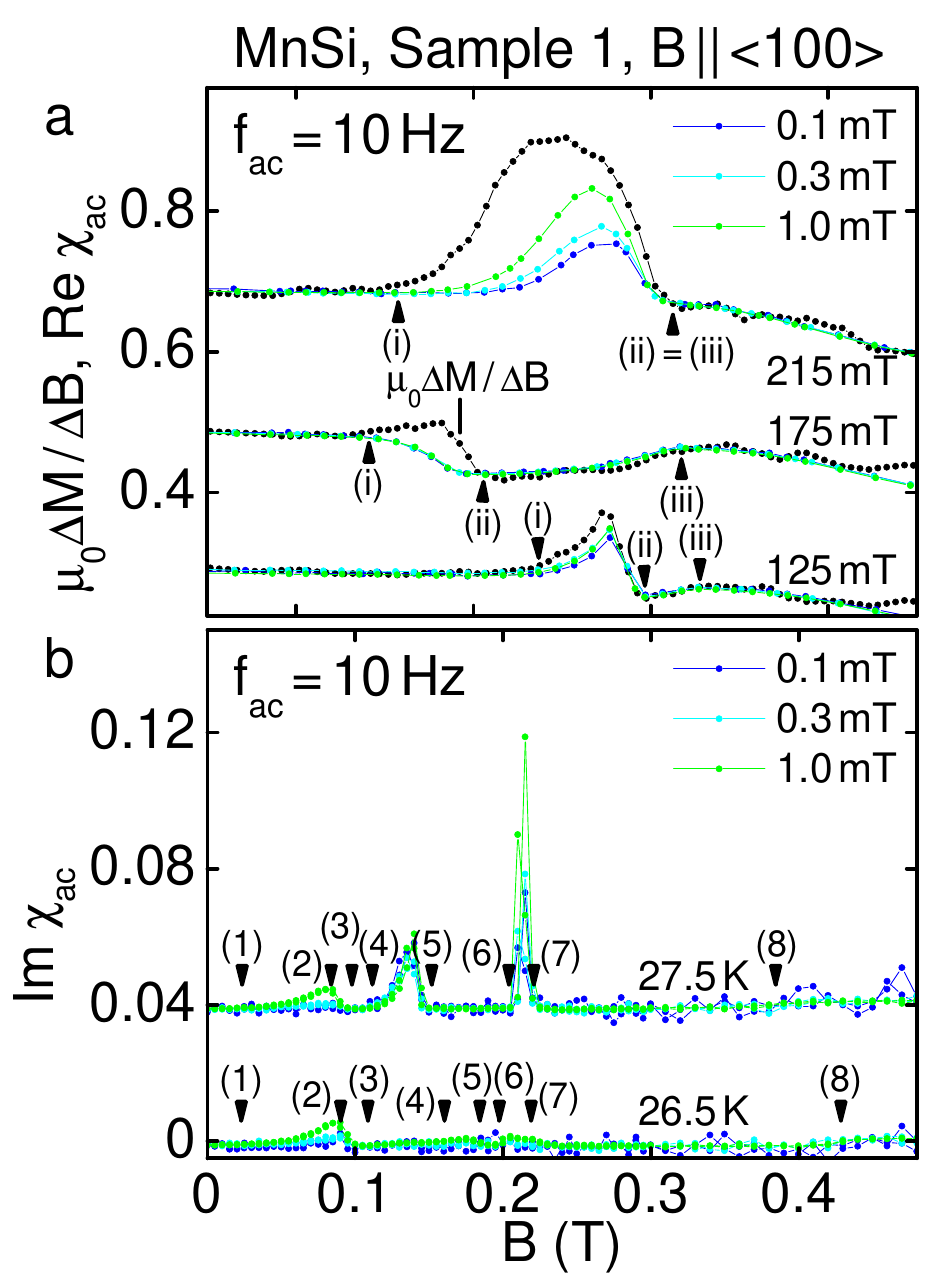}
\caption{(Color online) Typical temperature dependence of the susceptibility inferred from the magnetization, {\ddmddb}, and the real and imaginary parts of the ac susceptibility, {\rechi} and {\imchi}, respectively, at selected fields for various excitation amplitudes. Data were recorded for an excitation frequency of 10\,Hz. {\ddmddb} was calculated from magnetization data recorded for a field difference of 5\,mT. Curves at different magnetic fields have been shifted by constant values for clarity. See Table \ref{definitions-2} for a summary of the definitions of temperature values.}
\label{TSLayoutAmp}
\end{figure}

For lower magnetic fields the magnetization still displays a shallow minimum, while  the maximum in {\ddmddb} vanishes gradually and develops into a minimum for fields below 200\,mT as shown in Figs.\,\ref{TSLayoutSusVgl2}\,(b) through \ref{TSLayoutSusVgl2}\,(d). This corresponds still to trace (ii) in Fig.\,\ref{APhaseschematic}, where the two steps are now sufficiently separated. Finally a situation more akin to trace (iv) is shown in Fig.\,\ref{APhaseschematic}, where only one step is seen. We thereby continue to denote the lower and upper boundary of the feature in {\ddmddb} as $T_{\rm A1}^-$ and $T_{\rm A2}^+$, respectively. 

For even lower magnetic fields the magnetization develops a shallow maximum, as shown in Figs.\,\ref{TSLayoutSusVgl2}\,(e) and \ref{TSLayoutSusVgl2}\,(f). This corresponds to trace (iii) in Fig.\,\ref{APhaseschematic}, where the two steps are rather close to each other. In comparison,  {\ddmddb} develops again a maximum followed by a minimum, where the temperature of the maximum in the magnetization  coincides with the change of {\ddmddb} from a maximum to a minimum, i.e., the maximum in the magnetization is located in the center of the temperature range of the features in {\ddmddb}. 

The magnetic field dependence of {\rechi} does not track  {\ddmddb} in the regime of the A phase as shown in Fig.\,\ref{TSLayoutSusVgl2}. Here we define $T_{\rm A1}^+$ at the point where the difference between {\ddmddb} and {\rechi} vanishes. Typical data underscoring the importance of the excitation amplitude and frequency for {\rechi} in temperature sweeps are shown in Figs.\,\ref{TSLayoutFreq} and \ref{TSLayoutAmp}. For increasing frequency the features in {\rechi} become less pronounced and the peak in {\imchi} gets slightly larger [Figs.\,\ref{TSLayoutFreq}\,(a) and \ref{TSLayoutFreq}\,(b)]. Likewise {\rechi} and {\imchi} display a substantial dependence on the excitation amplitude, where the signal increases with increasing amplitude [Figs.\,\ref{TSLayoutAmp}\,(a) and \ref{TSLayoutAmp}\,(b)]. For all excitation amplitudes {\rechi} is smaller than {\ddmddb}. Hence, when trying to infer phase boundaries from the ac susceptibility measured at a single frequency or amplitude alone there is an inherent risk for misinterpretation.

\begin{figure}
\includegraphics[width=0.4\textwidth]{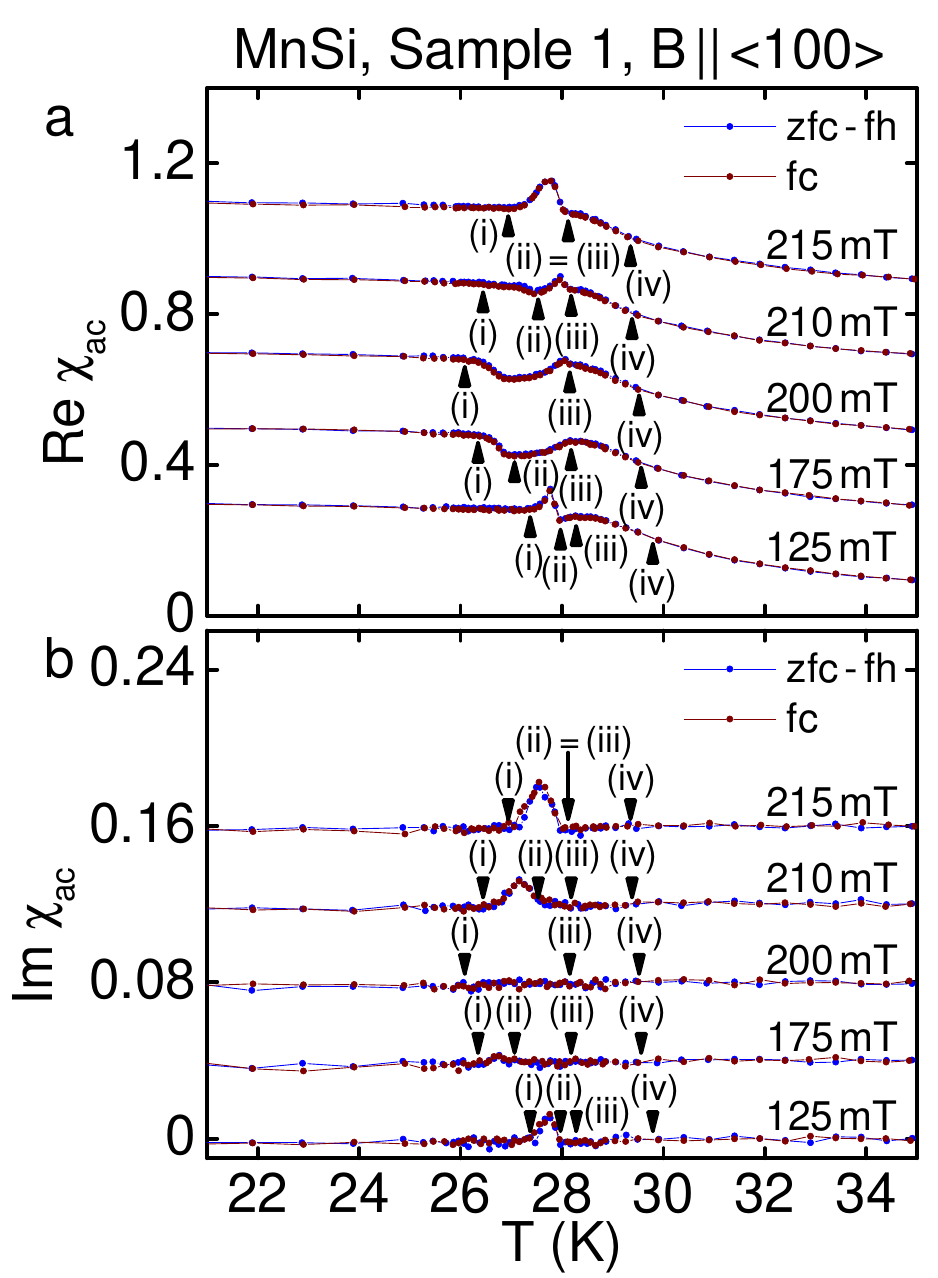}
\caption{(Color online) Comparison of the real and imaginary parts of the ac susceptibility, {\rechi} and {\imchi}, respectively, as recorded while field heating after zero-field cooling (zfc-fh) and while field cooling (fc). Data were recorded  in sample 1 for an excitation amplitude of 0.1\,mT along {\ozz} at 10\,Hz. Sets of curves for each field value are shifted by constants for clarity. No differences are observed in MnSi between zfc-fh and fc. This is contrasted by the properties of doped systems such as {\mfs}, {\mcs}, and {\fcs}, which display distinct differences as shown below.}
\label{TSLayoutZFCvsFC}
\end{figure}

As a final issue of the temperature dependence we have also explored the role of the field and temperature history. For this purpose we determined the difference between data recorded when heating the sample in an applied field after zero-field cooling (zfc-fh) and data recorded when cooling the sample after the field was applied at a temperature substantially higher than the helimagnetic transition temperature $T_c$ (fc). Here we find no differences whatsoever, as illustrated by typical data of {\rechi} and {\imchi} shown in Fig.\,\ref{TSLayoutZFCvsFC}. In contrast, all doped systems we have studied so far, notably {\mfs}, {\mcs}, and {\fcs},  display distinct differences between zfc-fh and fc (examples will be given below). In fact, an observation of differences between zfc-fh and fc in nominally stoichiometric systems with a complex metallurgy, say FeGe or MnGe, may hint at limited sample quality. 

In the spirit of a first summary we note that even though the temperature dependencies of $M$, {\ddmddb}, {\rechi}, and {\imchi} reported above at first sight might suggest a complex magnetic phase diagram in the regime of the A phase, the consistency of the temperature dependence with the magnetic field dependence unambiguously reveals that the phase diagram consists of a single pocket of the A phase. As an intrinsic property of the A phase the magnetization forms a tilted plateau with a slightly broadened first-order boundary. In fact, our temperature-dependent data are perfectly consistent with those reported by Kadowaki {\ea}, \cite{Kadowaki:JPSJ82} clearly establishing that the phase diagram proposed in Ref.\,\cite{Kadowaki:JPSJ82} is incorrect. The complexities in the temperature dependent data originate in the first-order boundary of the A phase where the A phase coexists in a small parameter range with the conical state. 

\section{Further issues}

The magnetization and ac susceptibility data reported in Secs. \ref{field} and \ref{temperature} clearly establish a single pocket of the A phase bounded by a regime of phase coexistence for field parallel to a crystallographic {\ozz} axis. This raises two questions. First, what does the phase diagram look like for field parallel to directions other than {\ozz}? Second, is the inconsistency between the magnetization and ac susceptibility at the helical to conical transition and at the border of the A phase extrinsic or intrinsic?

Motivated by these questions we now address changes of the properties reported above for samples with different shape and hence demagnetizing fields. We further address differences of the magnetization and the ac susceptibility, and thus the magnetic phase diagram, between magnetic field parallel to {\ozz}, {\ooz}, and {\ooo}. As a final aspect, we consider in further detail the frequency dependence of the ac susceptibility.

\begin{figure}
\includegraphics[width=0.4\textwidth]{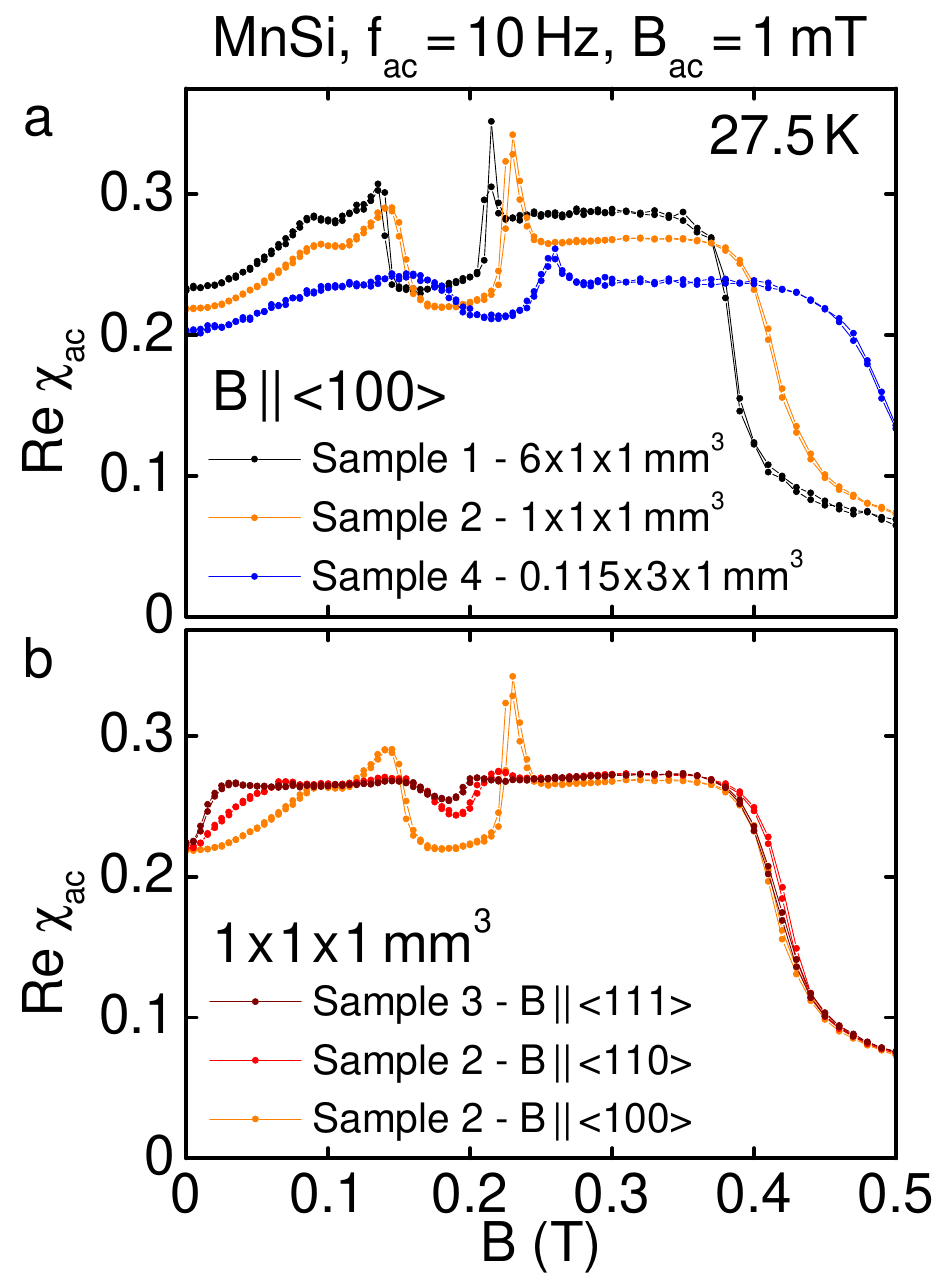}
\caption{(Color online) Typical ac susceptibility data for different sample shapes and orientations as a function of applied field. Data were recorded at an excitation amplitude of 1\,mT and 10\,Hz. (a) Real part of the ac susceptibility, {\rechi}, for three different sample geometries. The pronounced qualitative and quantitative differences for different sample shapes are purely due to demagnetizing fields. (b) Real part of the ac susceptibility, {\rechi} for a cubic sample shape and various crystallographic directions. Differences are purely due to differences of crystallographic orientation.}
\label{FSLayoutOverview}
\end{figure}

\subsection{Sample shape}

\begin{figure}
\includegraphics[width=0.4\textwidth]{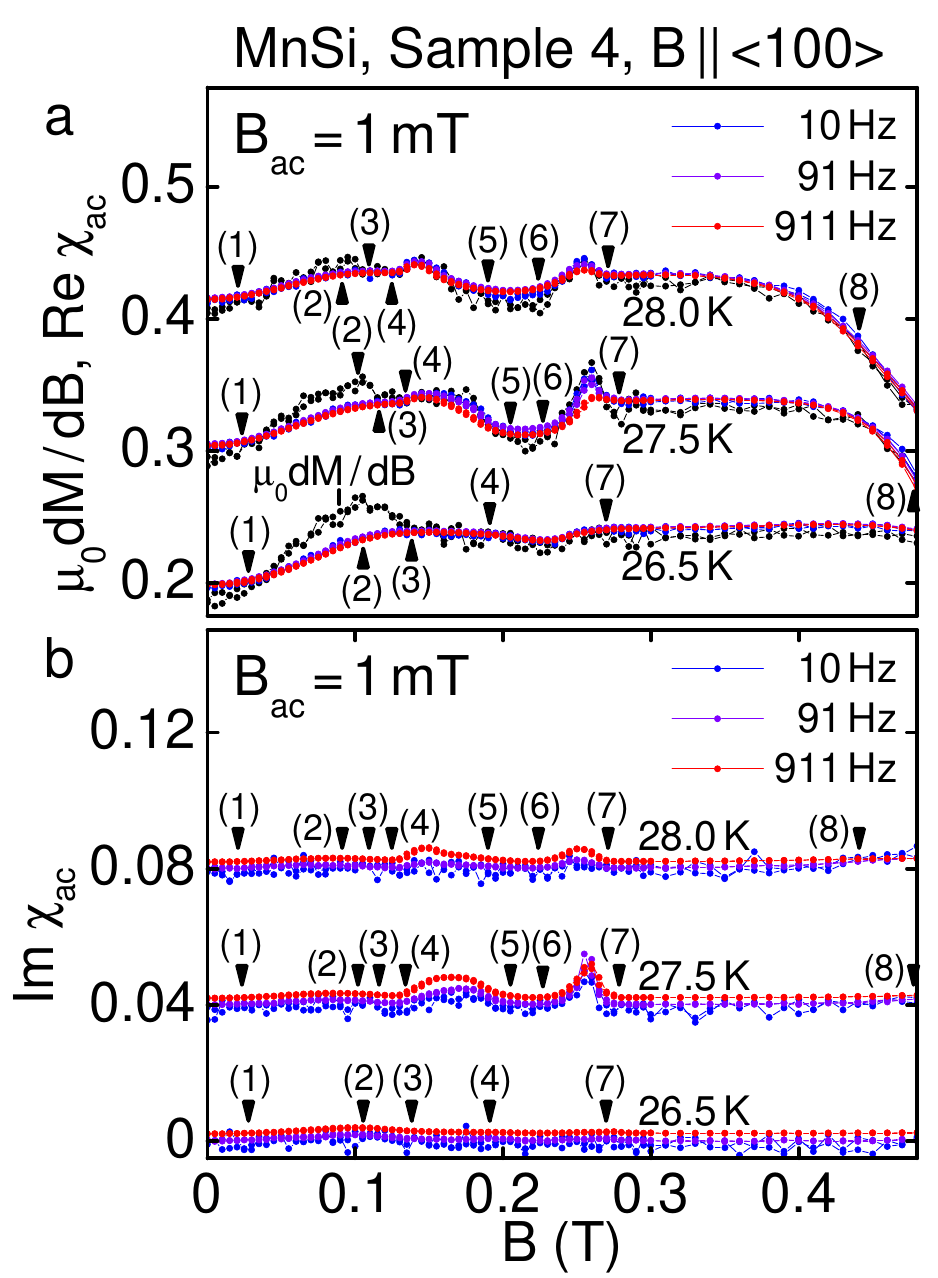}
\caption{(Color online) Typical field dependence of the susceptibility inferred from the magnetization, {\dmdb} or {\ddmddb}, respectively, and the real and imaginary parts of the ac susceptibility, {\rechi} and {\imchi}, respectively. Data shown here were recorded in sample 4 (a thin platelet) to track the role of large demagnetizing fields. The AC susceptibility data were recorded for an excitation amplitude of 1\,mT along {\ozz} at various excitation frequencies. Sets of curves for each temperature have been shifted by a constant for clarity. Values of the transition fields and temperatures have been determined in panel (a). The same values are marked in panel (b).}
\label{LayoutFreqAmpThinFS}
\end{figure}

\begin{figure*}
\includegraphics[width=0.95\textwidth]{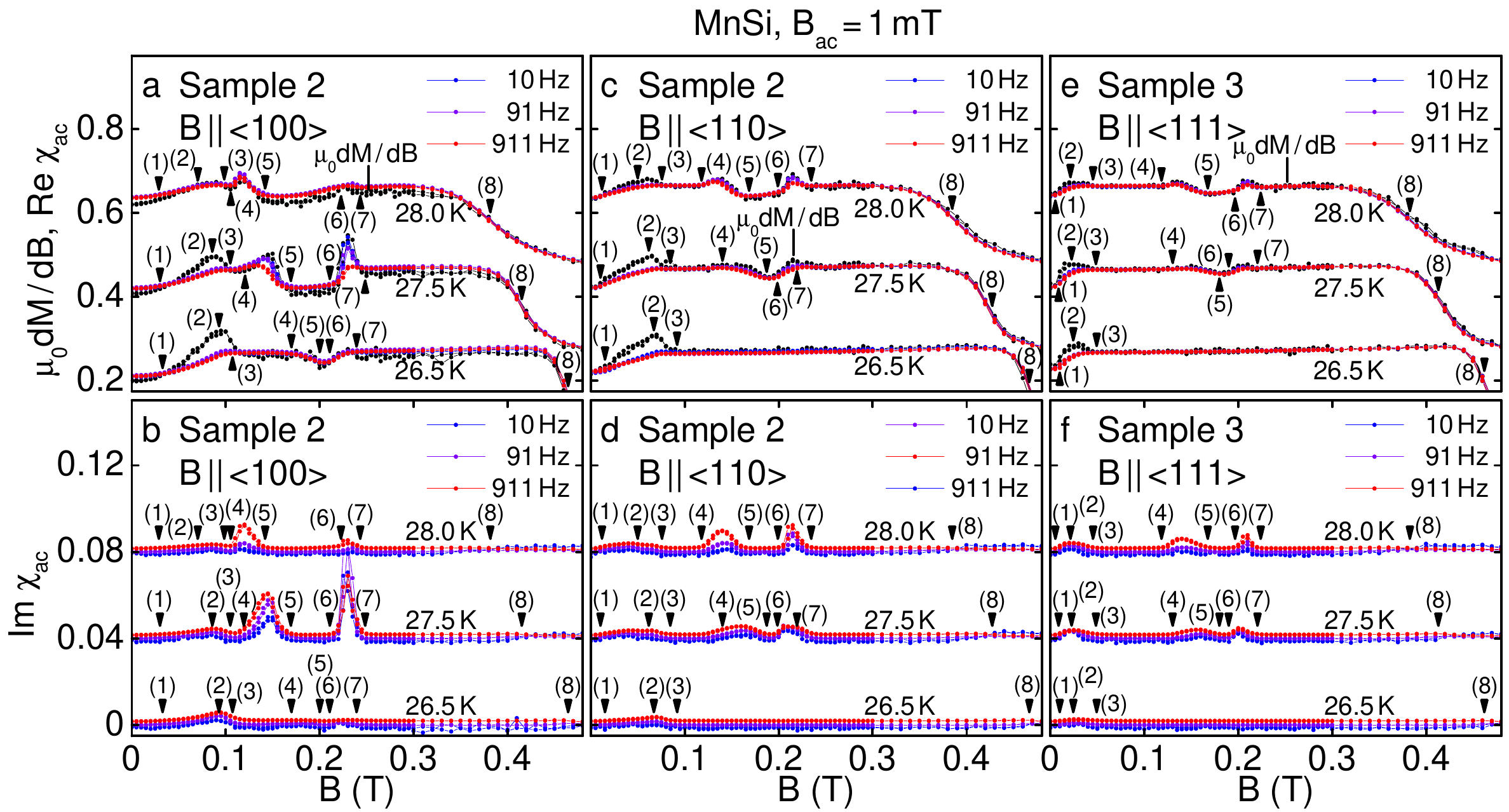}
\caption{(Color online) Comparison of the magnetic field dependence of the susceptibility calculated from the magnetization, {\dmdb}, and the real and imaginary parts of the ac susceptibility, {\rechi} and {\imchi}, respectively, at selected frequencies. Data shown here were recorded for magnetic field applied along $\langle100\rangle$, $\langle110\rangle$, and $\langle111\rangle$ in samples 2 and 3. Both samples are cubic for the field directions studied. Thus even though data are shown as a function of applied fields they may be readily compared. The ac susceptibility was measured for an excitation amplitude of 1\,mT. Sets of curves for the same temperature are shifted by constants for clarity.}
\label{FSLayoutFreqAmp100vs110vs111}
\end{figure*}

\begin{figure*}
\includegraphics[width=0.95\textwidth]{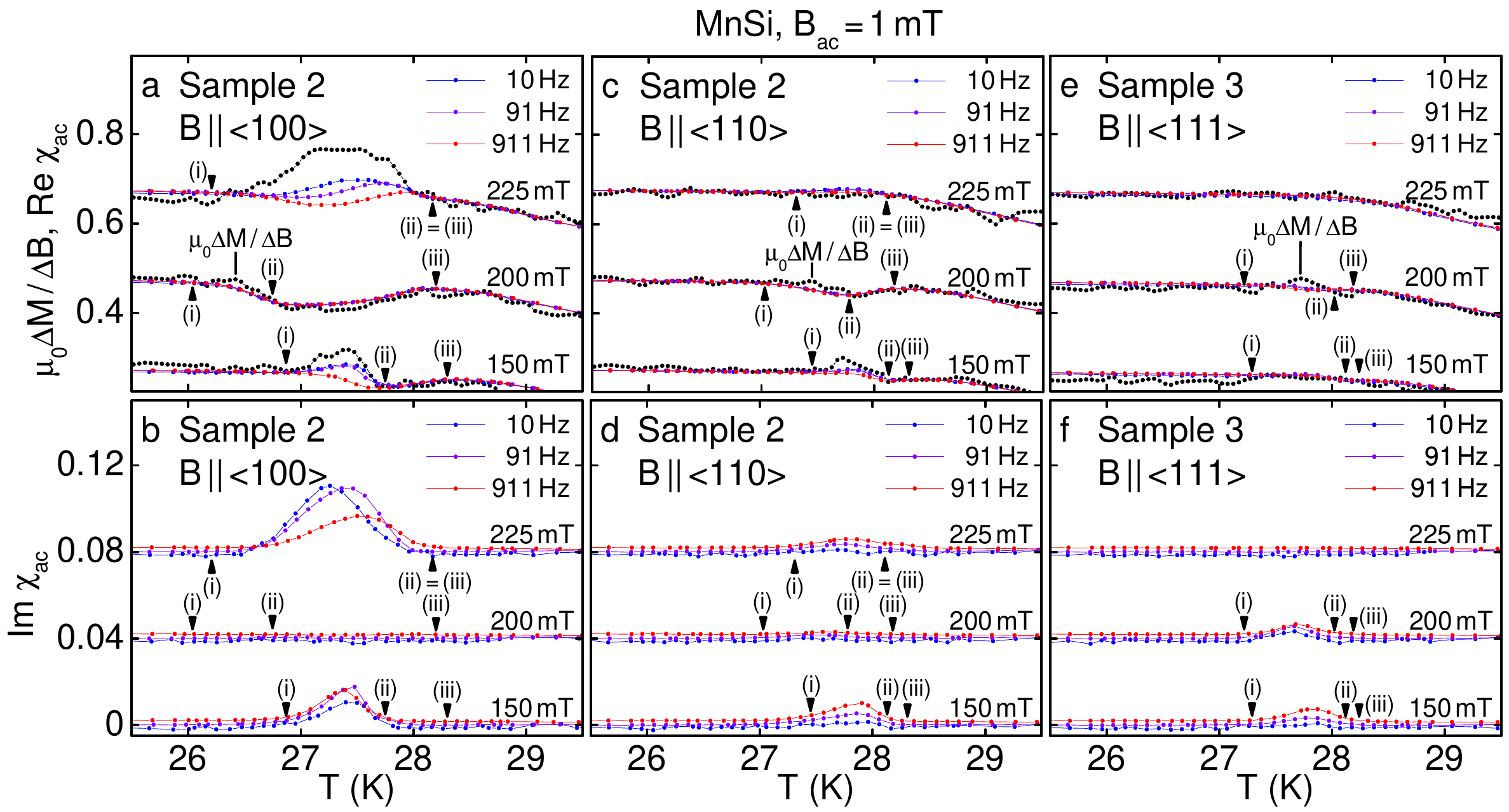}
\caption{(Color online) Comparison of the temperature dependence of the susceptibility calculated from the magnetization, {\ddmddb}, and the real and imaginary parts of the ac susceptibility, {\rechi} and {\imchi}, respectively, at selected frequencies. Data shown here were recorded for magnetic field applied along {\ozz}, {\ooz}, and {\ooo} in samples 2 and 3. Both samples are cubic for the field directions studied. Thus even though data are shown as a function of applied fields they may be readily compared. The ac susceptibility was measured for an excitation amplitude of 1\,mT. {\ddmddb} was calculated from magnetization data recorded for a field difference of 5\,mT. Sets of curves for the same magnetic field values are shifted by constants for clarity.}
\label{TSLayoutFreqAmp100vs110vs111}
\end{figure*}

Shown in Fig.\,\ref{FSLayoutOverview}\,(a) is the real part of the ac susceptibility {\rechi} as a function of applied magnetic field along the {\ozz} axis for three different sample shapes. Data were recorded for an excitation amplitude of 1\,mT and an excitation frequency of 10\,Hz. The sample shapes studied here were a bar, a cube, and a platelet denoted as samples 1, 2, and 4, respectively (cf. Table\,\ref{samples}). The samples shapes were chosen such that they differ substantially in terms of their demagnetizing factors. With increasing demagnetizing fields the characteristic transition fields are not only shifted to higher field values. More important perhaps, the signatures at the transition become less pronounced. Data sets recorded in sample 2 are shown in Figs.\,\ref{FSLayoutFreqAmp100vs110vs111} and \ref{TSLayoutFreqAmp100vs110vs111}. 
Data as a function of field recorded in sample 4, the thin platelet, are shown in Fig.\,\ref{LayoutFreqAmpThinFS} (the temperature dependent data are not shown).

\begin{figure}
\includegraphics[width=0.48\textwidth]{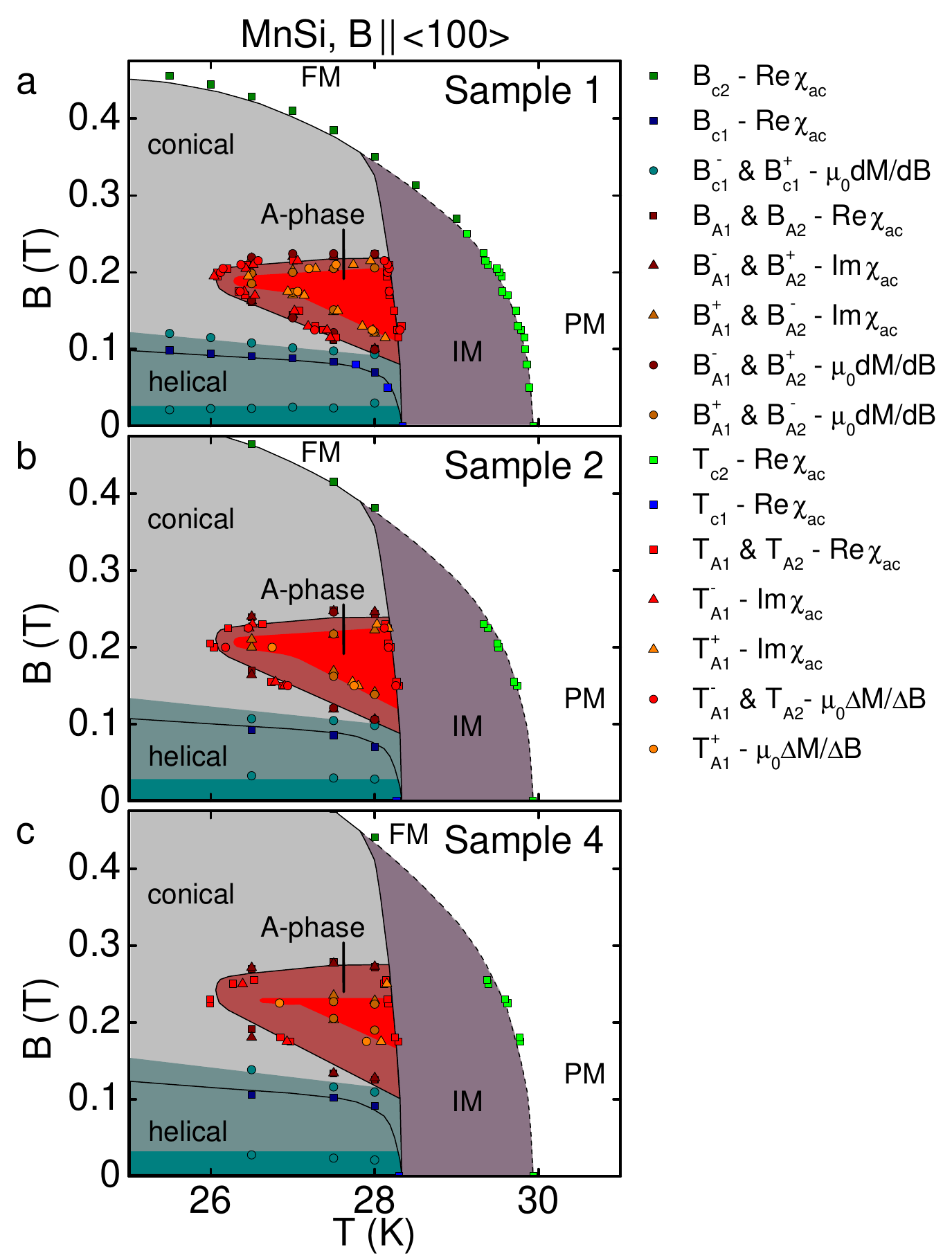}
\caption{(Color online) Magnetic phase diagram of MnSi as a function applied magnetic field along {\ozz} for different sample shapes. Sample 1 is a bar oriented along the applied field; sample 2 is a cube and sample 4 a thin platelet oriented perpendicular to the applied field. With increasing demagnetization fields the anomalies at $H_{\rm c1}$ and for the A phase broaden considerably. Data points between the regime marked IM and the paramagnetic state at higher temperatures represent a crossover.}
\label{PDLayoutBext}
\end{figure}

Following accurately the definitions of the characteristic field and temperature values given above, the phase diagrams of samples 1, 2, and 4 for field along {\ozz} have been determined as a function of applied magnetic fields as shown in Fig.\,\ref{PDLayoutBext}. As a function of temperature the helimagnetic transition temperature as well as the overall extent of the A phase are unchanged. Not surprisingly, the magnetic field dependence with increasing demagnetization effects is shifted to higher fields. In contrast, the extent of the regime of inconsistency between magnetization and ac susceptibility at the boundary of the A phase increases at the expense of the phase-pure part of the A phase. The increase of the regime of the inconsistency is thereby stronger than expected due to the demagnetizing fields. This suggests that the inconsistency is not an intrinsic characteristic.

Shown in Fig.\,\ref{LayoutFreqAmpThinFS} is the temperature and field dependence of {\dmdb} and {\rechi} at various frequencies in the A phase of sample 4. Even though data values are rather small and therefore noisy, the same general trends may be observed as for the other samples. It is thereby also important to note that the demagnetizing effects in sample 4 are much stronger than for any of the other samples. In fact, we believe that they can no longer be corrected effectively by approximating the sample shape with an ellipsoid. 

\subsection{Crystallographic orientation}

In order to explore systematically the importance of the crystallographic orientation we performed measurements on two cubic samples, denoted as 2 and 3. Starting with sample 2 we performed first field and temperature sweeps for {\ozz} as shown in Figs.\,\ref{FSLayoutFreqAmp100vs110vs111}\,(a) and \ref{FSLayoutFreqAmp100vs110vs111}\,(b)  and \ref{TSLayoutFreqAmp100vs110vs111}\,(a) and \ref{TSLayoutFreqAmp100vs110vs111}\,(b), respectively. A comparison with sample 1 (the bar for which the field  was applied parallel  to the long axis), quickly establishes that both samples display the same qualitative features. The cube thereby exhibits a slight rounding of the maxima at the various magnetic transitions. In contrast, there are no additional features between sample 1 and 2 that might hint at an important role other than the usual changes of the internal magnetic field.

We next measured sample 2 for magnetic field applied parallel to the {\ooz} axis. Since the faces of the cube were oriented such that the sample shape was identical for the two field directions, this allowed us to obtain a direct comparison of qualitative and quantitative differences. Typical data are shown in Figs.\,\ref{FSLayoutFreqAmp100vs110vs111}\,(c) and \ref{FSLayoutFreqAmp100vs110vs111}\,(d) and \ref{TSLayoutFreqAmp100vs110vs111}\,(c) and \ref{TSLayoutFreqAmp100vs110vs111}\,(d), respectively. While the field and temperature dependencies are similar, the perhaps most important difference concerns the field and temperature values of the anomalies. Clearly, as a function of temperature the extent of the A phase is much more narrow than for {\ozz}, in perfect agreement with the literature.

We finally performed an analogous set of measurements for the cubic sample 3, where we studied the properties for field parallel to {\ooz} and {\ooo}. This way we first confirmed that sample 3 showed accurately the same field and temperature dependence for {\ooz}. Taken together this way we obtained data that can be compared directly. As shown in Fig.\,\ref{FSLayoutFreqAmp100vs110vs111} a major difference concerns the absolute value of the helical to conical transition. For field along {\ozz} the value of $B_{\rm c1}$ is largest as expected of the magnetically hard axis, while $B_{\rm c1}$ is lowest for the magnetically soft {\ooo} axis. This compares with the A phase, where the signatures are most pronounced for the {\ozz} axis, while they are weakest for the {\ooo} axis. 

The same trends are also seen in the temperature dependencies recorded for samples 2 and 3. For the {\ozz} axis the signature of the A phase is rather pronounced, while is it is rather shallow and the essentially the same for the {\ooz} and {\ooo} axes with {\ooo} being slightly weaker. We return to the resulting phase diagrams further below.

\subsection{Detailed frequency dependence}

\begin{figure}
\includegraphics[width=0.45\textwidth]{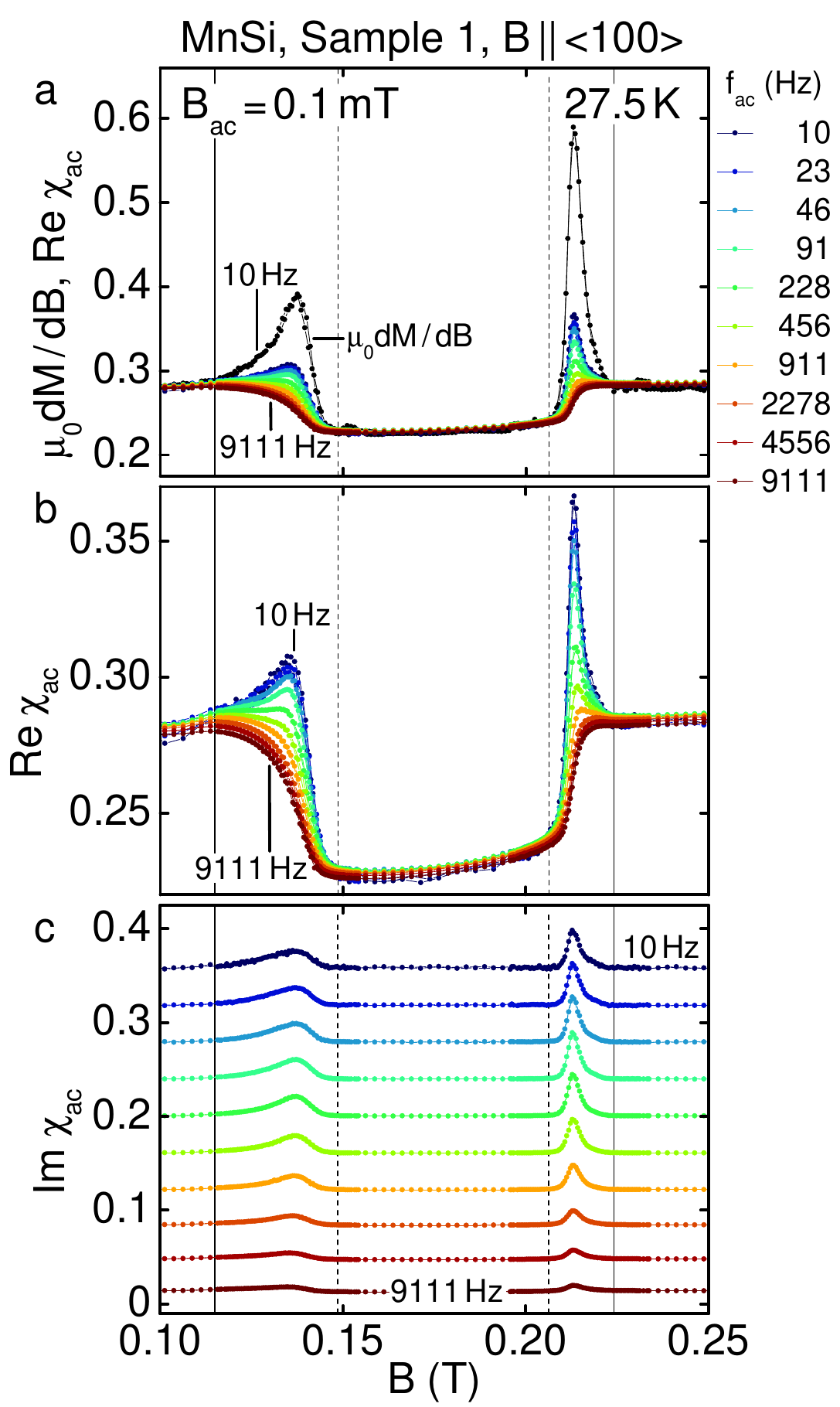}
\caption{(Color online) Comparison of the susceptibility calculated from the measured magnetization, {\dmdb}, and the ac susceptibility as measured at a wide range of different frequencies. The maxima at the lower and upper boundary of the A phase decrease dramatically with increasing frequency and vanish altogether for very large frequencies. Likewise maxima in the imaginary part of the susceptibility at the phase boundaries also decrease substantially with increasing frequency.}
\label{FSLayoutFreqDetail}
\end{figure}

It is finally instructive to determine the frequency dependence of the ac susceptibility in further detail. Shown in Fig.\,\ref{FSLayoutFreqDetail} is {\rechi} in a field range crossing the A phase for a temperature of 27.5\,K. Increasing the excitation frequency from 10\,Hz to 10\,kHz the maxima at the transition between the conical state and the A phase decrease and vanish.  In this frequency range {\rechi} is always smaller than {\dmdb}. A similar frequency dependence is also seen in {\imchi}, which displays a finite contribution in the range of the transition that decreases with increasing frequency [Fig.\,\ref{FSLayoutFreqDetail}\,(b)]. 

\begin{figure}
\includegraphics[width=0.45\textwidth]{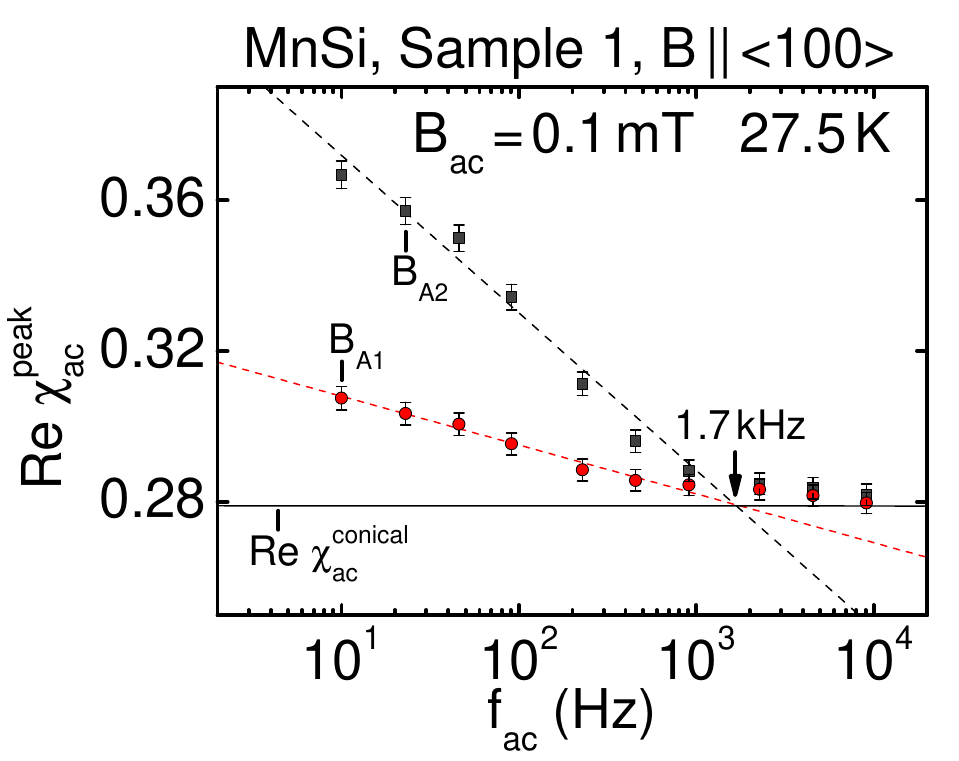}
\caption{(Color online) Frequency dependence of the height of the peak in the ac susceptibility at the lower and upper boundary of the A phase denoted as $B_{A1}$ and $B_{A2}$, respectively. Both maxima decrease exponentially and vanish for frequencies above $\sim1.7\,{\rm kHz}$ and reach the size of the susceptibility in the conical phase.}
\label{FSLayoutChiMaxFreq}
\end{figure}

Shown in Fig.\,\ref{FSLayoutChiMaxFreq} are the peak magnitudes of {\rechi} at the transition to the A phase. On a logarithmic frequency scale the maxima at $B_{\rm A1}$ and $B_{\rm A2}$ both decrease linearly and vanish around the same characteristic frequency $f_{\rm A}\sim1.7\,{\rm kHz}$. For frequencies exceeding $f_{\rm A}$ the susceptibility at the transition approaches the value of the conical phase. When extrapolating the frequency dependence towards low frequencies the peak height observed in {\dmdb} would hypothetically only be reached for an excitation frequency around $10^{-5}\,{\rm Hz}$. 

In contrast to the transition between the conical phase and A phase we do not observe any evidence for a maximum in {\rechi} at the border between the helical and the conical state in the frequency range between 10\,Hz and 10\,kHz that we have studied. This implies that the characteristic frequency $f_{\rm c1}$ must be below 10\,Hz; i.e., $f_{\rm c1}\ll f_{\rm A}$. This is consistent with spin torque experiments in the skyrmion lattice phase of MnSi, which demonstrate a very weak pinning of the skyrmion lattice to defects and disorder. \cite{Jonietz:Science2010,Schulz:NaturePhysics2012} It is also consistent with the excellent long-range nature of the skyrmion lattice phase seen recently in small-angle neutron scattering. \cite{Adams:PRL2011} The small ac excitation may therefore be much more efficient to generate a response of domains of the A phase at the border to the conical phase as compared with domains at the helical to conical transition.

\subsection{Relevance for other B20 compounds}

In this section we consider the question to what extent our observations are evidence of a universal magnetic phase diagram of helimagnetic B20 compounds. In turn this concerns foremost the disagreement between the magnetization and the ac susceptibility that we observe at the helical to conical transition and at the border of the A phase in pure MnSi as a generic property of all helimagnetic B20 transition metal compounds. In our study we have addressed this question in two ways, namely by considering the importance of disorder in MnSi when doping it with Fe and, second, by considering the properties of a rather different system -- the heavily doped semiconductor {\fcs}. As a final aspect we discuss the results of ac susceptibility measurements across the magnetic phase diagram of FeGe. \cite{Wilhelm:PRL2011}

To study the importance of disorder we have measured simultaneously the ac susceptibility and the magnetization of {\mfs} ($x=0.04$). Previous studies have established that substitutional doping of MnSi with Fe or Co causes a suppression of the helimagnetic transition temperature, driving the system through several quantum phase transitions. \cite{Bauer:PRB2010,Manyala:NatureMaterials2004} In {\mfs} with $x=0.04$ the helimagnetic transition temperature is suppressed to $T_ c=15\,{\rm K}$. The magnetization, the ac susceptibility as measured at 911\,Hz, and the specific heat of the same sample studied here were previously reported in Ref.\,\cite{Bauer:PRB2010}. This paper reports also the magnetic phase diagram, which is qualitatively reminiscent of pure MnSi. 

\begin{figure}
\includegraphics[width=0.4\textwidth]{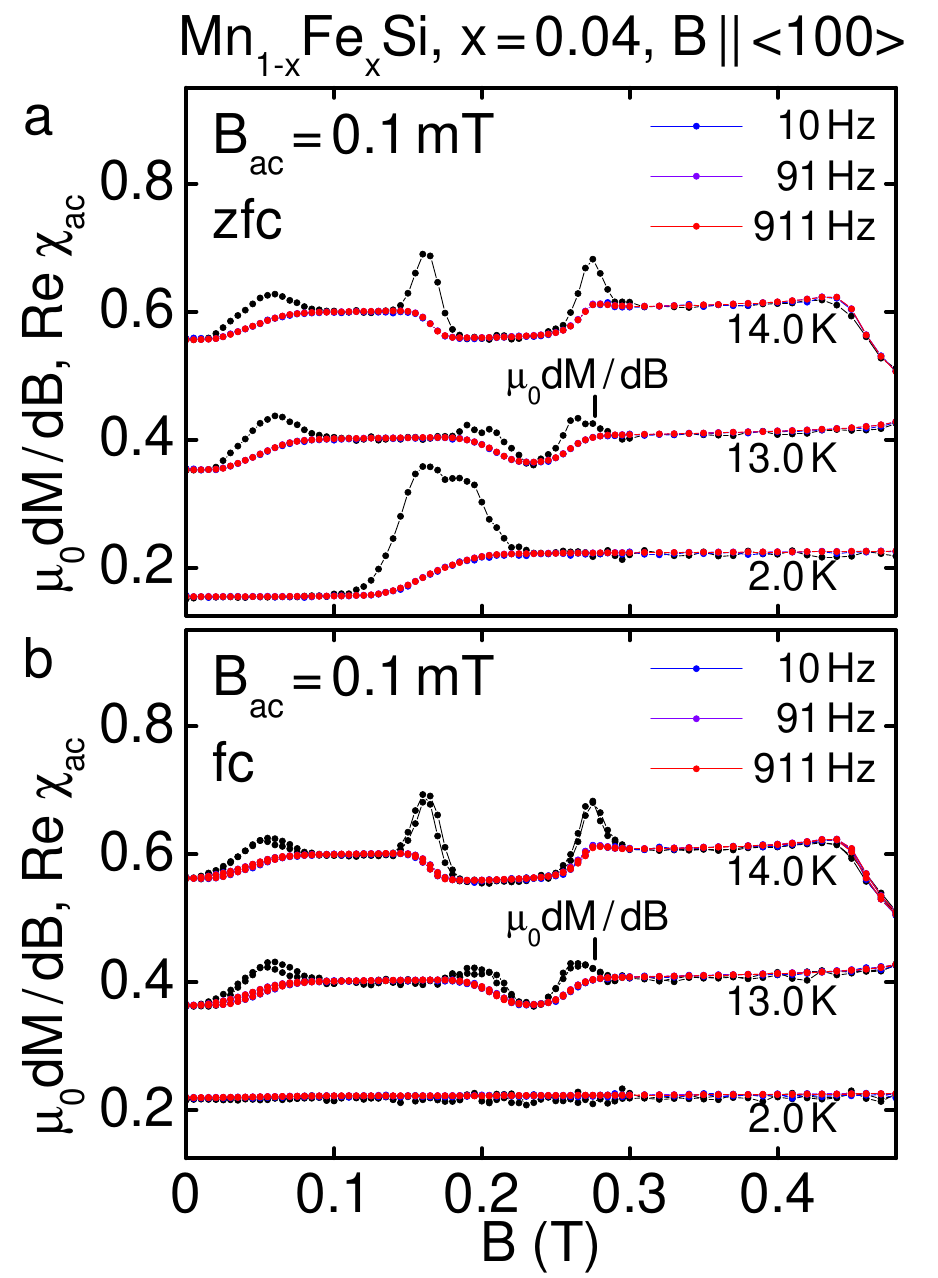}
\caption{(Color online) Comparison of the susceptibility {\dmdb} calculated from the magnetization with the ac susceptibility for {\mfs} ($x=0.04$). (a) Data after zero-field cooling. (b) Data after field cooling. The same qualitative differences are observed as in MnSi with additional differences between the zfc and fc data in the helical state.}
\label{LayoutMnFeSi004}
\end{figure}

Shown in Fig.\,\ref{LayoutMnFeSi004} is a comparison of the susceptibility calculated from the magnetization, {\dmdb}, and the ac susceptibility for field along {\ozz}. For the purpose of the study reported here field sweeps at three different temperatures were carried out, one well below $T_c$ at 2\,K, the other two at 13\,K and 14\,K at the upper and lower end of the A phase, respectively. The ac susceptibility was thereby recorded for three different frequencies. 

Data shown in Fig.\,\ref{LayoutMnFeSi004}\,(a) were obtained after zero-field cooling. Qualitatively these data are highly reminiscent of pure MnSi, with the same general features at the helical to conical transition at $B_{c1}$ and at the border of the A phase. In particular we observe again a clear deviation between {\dmdb}, which displays distinct maxima, and the ac susceptibility which does not display maxima. However, in contrast to pure MnSi, we do not observe any difference of the ac susceptibility between 10\,Hz and 911\,Hz, the lowest and highest frequency measured, over the entire field range studied. This suggests that the disorder in slightly Fe-doped {\mfs} shifts the frequency dependence observed in MnSi at the border of the A phase to frequencies well below 10\,Hz. 

The second important difference between pure MnSi and {\mfs} with $x=0.04$ concerns the transition between the helical and the conical phase. After zero-field cooling $B_{c1}$ displays a clear temperature dependence and increases with decreasing temperature. \cite{Bauer:PRB2010} In contrast, $B_{c1}$ is essentially temperature independent in MnSi for temperatures well below $T_c$. As pointed out before, this suggests that disorder affects the transition at $B_{c1}$ consistent with a reorientation and defect-related pinning. Moreover, as also pointed out before, once field cooled, the conical phase in {\mfs} ($x=0.04$) is stable for zero magnetic field and the system does not return to the helical state. This conjecture was so far based on ac susceptibility as measured at 911\,Hz only. The magnetic field dependence of the field-cooled state confirms these conclusions, as shown in Fig.\,\ref{LayoutMnFeSi004}\,(b). While the anomalies and the difference between {\dmdb} and {\rechi} in the vicinity of $T_c$ are strongly reminiscent of the zfc state, there is no evidence for a helical to conical phase transition at 2\,K. The reappearance of the minima at low fields for higher temperatures may be attributed to the effects of thermal excitations into the helical state.

\begin{figure}
\includegraphics[width=0.4\textwidth]{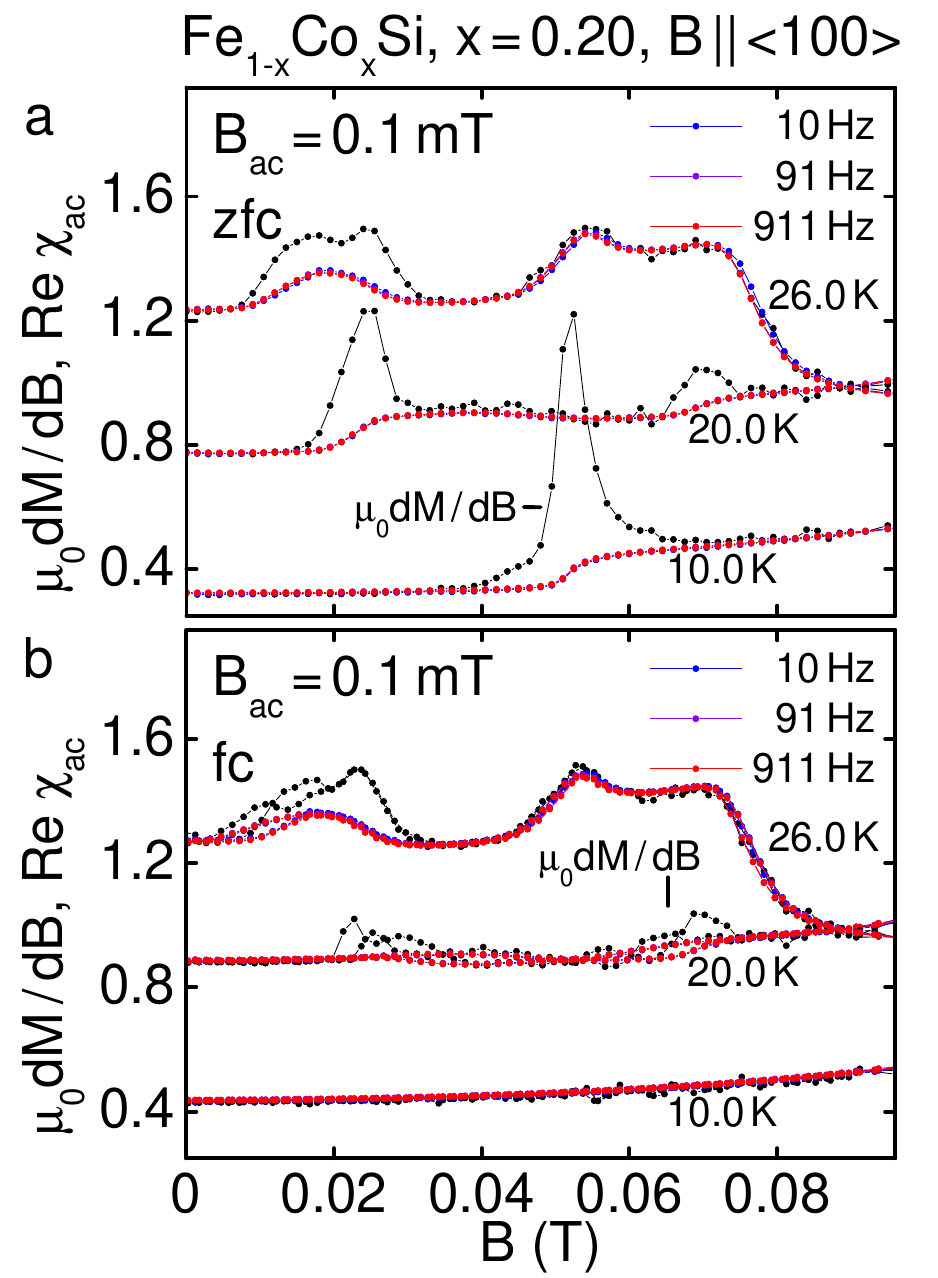}
\caption{(Color online) Comparison of the susceptibility calculated from the magnetization with the ac susceptibility for {\fcs} ($x=0.20$). (a) Data after zero-field cooling. (b) Data after field cooling. The same qualitative differences are observed as in MnSi with additional differences between zfc and fc data.}
\label{LayoutFeCoSi020}
\end{figure}

We turn next to the properties of {\fcs}, which differs from MnSi and {\mfs}\,($x=0.04$) in two important ways. First the amount of substitutional doping is much higher, and, second, {\fcs} is a heavily doped semiconductor instead of a metal. Nevertheless, the susceptibility calculated from the magnetization, {\dmdb}, and the ac susceptibility are qualitatively reminiscent of {\mfs}\,($x=0.04$) as shown in Fig.\,\ref{LayoutFeCoSi020}. Again there is no frequency dependence of the ac susceptibility.

Our results shed an unexpected light on a recent study of the B20 compound FeGe \cite{Wilhelm:PRL2011} in which the magnetic phase diagram was inferred from the ac susceptibility alone.  Measurements of the ac susceptibility were carried out at 1\,kHz in a PPMS, i.e., an apparatus identical to that used in our study. Moreover, a single crystal was studied grown by vapor transport with a rather poor residual resistivity ratio around 10 \cite{Pedrazzini:PRL2007}. The shape of the single crystal was ill defined making corrections of demagnetizing fields difficult. Keeping these constraints in mind a complex phase diagram was inferred from the ac susceptibility, suggesting several phase pockets and putative evidence for mesophase formation without any microscopic evidence whatsoever.

However, the qualitative field and temperature dependence of the ac susceptibility of FeGe (recorded at 1\,kHz) is strongly reminiscent of that reported here for MnSi and related compounds. This suggests that the phase diagram reported in Ref.\,\cite{Wilhelm:PRL2011} is not correct. For a complete phase diagram simultaneous measurements of the magnetization and ac susceptibility are necessary. Given the poor residual resistivity ratio we would thereby expect that the ac susceptibility does not show a frequency dependence for the same reasons that there is no frequency dependence in {\mfs}\,($x=0.04$) and {\fcs}\,($x=0.20$). Moreover, as a vapor transport grown sample there may be tiny variations of the composition, capable of introducing small local strains that might appear as several phase pockets.

\section{Conclusions
\label{conclusions}}

\begin{figure}
\includegraphics[width=0.48\textwidth]{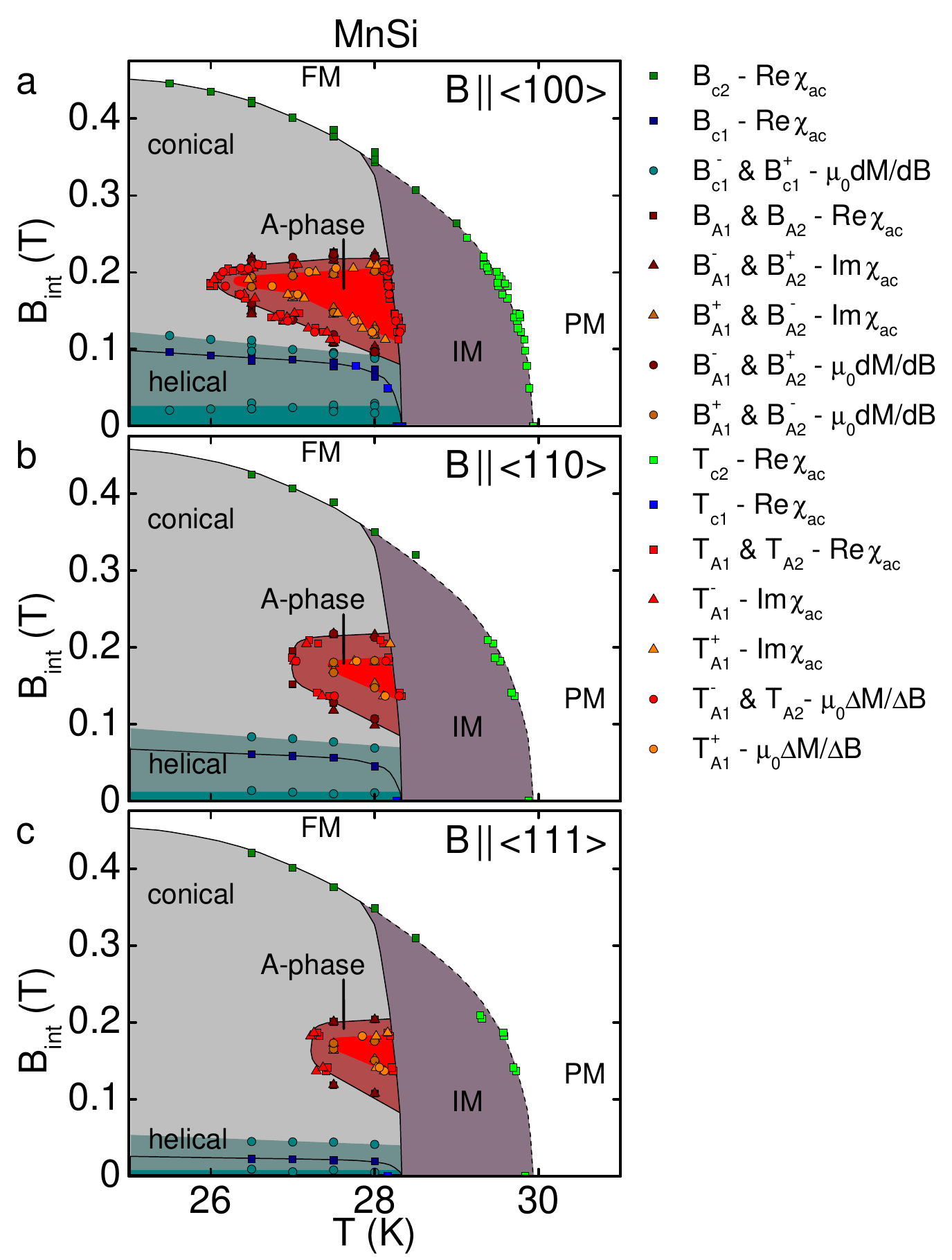}
\caption{(Color online) Magnetic phase diagram of MnSi for field along different crystallographic directions. Phase diagrams are shown as a function of internal field; the effects of demagnetizing fields have been corrected. Data shown here were inferred from samples 1 through 4. Data points between the regime marked IM and the paramagnetic state at higher temperatures represent a cross-over.}
\label{PDLayoutBint}
\end{figure}

We conclude this paper with the magnetic phase diagram of MnSi as a function of internal field shown in Fig.\,\ref{PDLayoutBint}. The phase diagrams shown here are based on the magnetization and ac susceptibility data recorded in samples 1, 2, and 3. Only the outer boundary of the A phase determined in 
sample 4 was included in Fig.\,\ref{PDLayoutBint}\,(a), since the width of the transition surrounding the A phase in sample 4 is much larger than seen in samples 1, 2, and 3. We attribute this additional broadening to demagnetizing fields that are more complex than can be corrected with the approximation as a rectangular prism. \cite{Aharoni:JAP1998}

The phase diagrams shown in Fig.\,\ref{PDLayoutBint} highlight several key properties. Namely, the extent of the A phase is largest for field parallel to {\ozz} and smallest for {\ooo}. This agrees with a previous report. \cite{Lamago:PhysicaB2006} Moreover, the lower critical field $B_{\rm c1}$ is largest for field parallel to {\ozz} and smallest for {\ooo}. Both the anisotropy of the A phase as well as $B_{c1}$ are due the same leading order magnetic anisotropy, which is fourth order in spin-orbit coupling. \cite{Bak:JPC1980,Nakanishi:SSC1980} The same fourth-order anisotropy term causes the zero-field helical order to propagate along {\ooo}. The temperature dependence of the upper critical field $B_{\rm c2}$ is the same for all three directions independent of the magnetic anisotropy within the accuracy of our measurements. 

Our study reveals also a potential source of errors in the determination of the magnetic phase transitions in helimagnetic B20 compounds. At the phase boundary between the helical and the conical phase, as well as the conical and the A phase, we observe rather pronounced differences between the susceptibility calculated from the magnetization, {\dmdb}, and the ac susceptibility, {\rechi}. General thermodynamic considerations and the imaginary part of the ac susceptibility, {\imchi}, identify the maxima in {\dmdb} as the response of large magnetic objects at very slow time-scales, equally consistent with a reorientation or weak first-order transition. Since this behavior occurs between the helical and the conical phase \textit{and} the conical and the A phase, there is no reason to believe that it represents evidence for spin textures with nontrivial topology. 

The disagreement between {\dmdb} and {\rechi} finally highlights the risk to determine phase transition lines from the ac susceptibility alone. In fact, the change of the isothermal magnetization inferred from the ac susceptibility between two different field values in the conical phase below and above the A phase would be inconsistent for a path across the A phase and a path surrounding the A phase. This by itself questions strongly the phase diagram of FeGe reported in Ref.\,\cite{Wilhelm:PRL2011}.  Moreover, in our simultaneous measurements of the magnetization and ac susceptibility we reproduce the magnetization reported by Kadowaki {\ea}. \cite{Kadowaki:JPSJ82} Here our data demonstrate unambiguously that the phase diagram reported in Ref.\,\cite{Kadowaki:JPSJ82} is not correct. Instead we find that the phase diagram of MnSi observed in the magnetization and ac susceptibility yields a single pocket of the A phase, regardless of crystallographic orientation, sample shape, excitation frequency, and excitation amplitude of the ac susceptibility. 

We finally note that none of the rather large number of microscopic measurements of the A phase reported to date hints at a more complex magnetic phase diagram than reported here. Moreover, the phase boundaries observed in microscopic studies, notably SANS, agree within experimental accuracy with the phase boundaries we infer from the magnetization and ac susceptibility. In turn this questions the evidence for an exotic mechanism at the heart of the skyrmion lattice phase. \cite{Leonov:preprint2010} In conclusion, the combination of excellent theoretical understanding of the A phase \cite{Muehlbauer:Science2009,Han:PRB2010} with the similarity of our data in MnSi, {\mfs}, and {\fcs} suggests that the phase diagram of MnSi represents a universal property of all B20 transition metal compounds with helimagnetic order. 

\textit{Note added in proofs.} We have confirmed that the same differences between the magnetization and ac susceptibility reported here exist also in the magnetic phase diagram of the multiferroic insulator Cu$_{2}$OSeO$_{3}$ (cf. Ref. \cite{Adams:PRL2012}). These results will be published elsewhere.

\acknowledgments
We gratefully acknowledge discussions with M. Garst and A. Rosch. We wish to thank M. Baenitz and M. Schmidt for discussions on their measurements of FeGe. Financial support through DFG TRR80 (From Electronic Correlations to Functionality), DFG FOR960 (Quantum Phase Transitions), and ERC AdG (291079, TOPFIT) is gratefully acknowledged. A.B. acknowledges financial support through the TUM graduate school.


\end{document}